\documentclass[pra,aps,showpacs,amssymb,amsmath,amstext,amsfont,noshowkeys,groupedaddress,12pt,nofootinbib]{revtex4-1}
\usepackage{graphicx}
\usepackage{accents}

\newcommand\zero[1]{\accentset{(0)}{#1}}

\newcommand\emm[1]{\accentset{(\mathbf{m})}{#1}}

\begin{document}

\setlength{\baselineskip}{17pt}
\numberwithin{equation}{section}
\newtheorem{theorem}{Theorem}[section]
\newtheorem{lemma}{Lemma}[section]
\newtheorem{corollary}{Corollary}[section]
\newtheorem{definition}{Definition}[section]
\newtheorem{example}{Example}[section]
\newtheorem{exercise}{Exercise}[section]
\newtheorem{notation}{Notation}[section]
\newtheorem{proposition}{Proposition}[section]
\setlength{\oddsidemargin}{0cm} \setlength{\evensidemargin}{0cm}
\setlength{\textheight}{20cm} \setlength{\textwidth}{16cm}
\newcommand{\PP}{\mathcal{P}}
\newcommand{\Gll}{\mathsf{Gl}}
\newcommand{\D}{\mathsf{D}}
\newcommand{\sph}{\mathsf{S}}
\newcommand{\R}{\mathbb{R}}
\newcommand{\I}{\mathbb{I}}
\newcommand{\Z}{\mathbb{Z}}
\newcommand{\N}{\mathbb{N}}
\newcommand{\C}{\mathbb{C}}
\newcommand{\B}{\mathbb{B}}
\newcommand{\Q}{\mathbb{Q}}
\newcommand{\T}{\mathbb{T}}
\newcommand{\cw}{\mathfrak{CW}}
\newcommand{\ab}{\alpha\beta}
\newcommand{\bg}{\beta\gamma}
\newcommand{\abg}{\alpha\beta\gamma}
\newcommand{\ag}{\alpha\gamma}
\newcommand{\ga}{\gamma\alpha}
\newcommand{\weak}{\text{weak}}
\newcommand{\dvol}{\text{dvol}}
\newcommand{\dd}{\text{d}}
\newcommand{\cc}{\mathfrak{c}}
\newcommand{\Or}{\mathbf{O}}
\newcommand{\SO}{\mathbf{SO}}
\newcommand{\U}{\mathbf{U}}
\newcommand{\SU}{\mathbf{SU}}
\newcommand{\Sp}{\mathbf{Sp}}
\newcommand{\Ad}{\text{Ad}}
\newcommand{\Aut}{\text{Aut}}
\newcommand{\strong}{\text{strong}}
\newcommand{\Dirac}{\text{Dirac}}
\newcommand{\Gl}{\text{Gl}}
\newcommand{\HH}{\mathbb{H}}
\newcommand{\YMe}{\mathcal{YM}_{\epsilon}}
\newcommand{\YM}{\mathcal{YM}}
\newcommand{\q}{\mathsf{q}}
\newcommand{\A}{\mathsf{a}}
\newcommand{\loc}{\text{loc}}
\newcommand{\ttilde}{\tilde{\intercal}}
\newcommand{\tperp}{\tilde{\perp}}
\newcommand{\Je}{\mathcal{J}_{\epsilon}}
\newcommand{\tr}{\mathsf{Tr}}
\newcommand{\so}{\mathfrak{so}}
\newcommand{\cat}{\text{cat}}
\newcommand{\Asf}{\mathsf{A}}
\newcommand{\hsf}{\mathsf{h}}
\newcommand{\Sum}{\sum_{i=1}^{n}}
\newcommand{\Ser}{\sum_{i=1}^{\infty}}
\newcommand{\p}{\partial}
\newcommand{\be}{\begin{equation}}
\newcommand{\ee}{\end{equation}}
\newcommand{\ba}{\begin{align}}
\newcommand{\bt}{\begin{theorem}}
\newcommand{\et}{\end{theorem}}
\newcommand{\bc}{\begin{corollary}}
\newcommand{\ec}{\end{corollary}}
\newcommand{\ed}{\end{document}}
\newcommand{\x}{(x^1, ..., x^n)}
\newcommand{\0}{\ ^{(0)}}
\newcommand{\1}{\ ^{(1)}}
\newcommand{\2}{\ ^{(2)}}
\newcommand{\z}{_{(0)}}
\newcommand{\one}{_{(1)}}
\newcommand{\two}{_{(2)}}
\newcommand{\ke}{_{(f)}}
\newcommand{\nn}{_{(n)}}
\newcommand{\el}{_{(l)}}
\newcommand{\ione}{(-\infty, 0]\times\R}
\newcommand{\onest}{{\R^-\times\R}}
\newcommand{\inst}{(-\infty, 0]\times\R^n}
\newcommand{\threest}{{\R^-\times\R^2}}
\newcommand{\st}{{\R^-\times\R^3}}
\newcommand{\nst}{{\R^-\times\R^n}}
\newcommand{\cst}{{(-\infty, 0]\times\R^3}}
\newcommand{\cnst}{{(-\infty, 0]\times\R^n}}
\newcommand{\bst}{{\partial (\R^-\times\R^3)}}
\newcommand{\tzero}{{\{0\}\times \R^3}}
\newcommand{\threetzero}{{\{0\}\times \R^2}}
\newcommand{\bnst}{{\partial (\R^-\times\R^n)}}
\newcommand{\ntzero}{{\{0\}\times \R^n}}
 \newcommand{\rinfty}{{\{\Vert  {\bf x}\Vert = R\,,\; x^0\leq 0\}}}
\newcommand{\grinfty}{{\{\Vert  {\bf x}\Vert \geq R\,,\; x^0\leq 0\}}}
\newcommand{\rinftyi}{{\{\Vert  {\bf x}\Vert = R_i\,,\; x^0\leq 0\}}}

\newcommand{\rinftyz}{{\{\Vert  {\bf x}\Vert = R\,,\; x^0= 0\}}}

\newcommand{\rinftyiz}{{\{\Vert  {\bf x}\Vert = R_i\,,\; x^0= 0\}}}

\newcommand{\ie}{{\mathcal I}_{es}}

\newcommand{\h}{\mathcal H}
\newcommand{\kk}{\mathcal K}
\newcommand{\bb}{\mathcal B } 
\newcommand{\ma}{\mathcal A}
\newcommand{\me}{\mathcal E}
\newcommand{\pp}{\mathcal P}

\newcommand{\uar}{\mathcal U_{{\bf a}; R}}
\newcommand{\uarone}{\mathcal U_{{\bf a}; R_1}}
\newcommand{\ur}{\mathcal U_R}
\newcommand{\ffi}{\Phi_ \varphi}
\newcommand{\fio}{{\Phi}_ {{\varphi}_0}}
\newcommand{\fiL}{{\Phi}_ {\varphi}^L}
\newcommand{\fioL}{{\Phi}_ {{\varphi}_0}^L}

\newcommand{\qed}{\hfill$\Box$ \medskip}
\newcommand{\cf}{\emph{cf.\;}}
\newcommand{\sn}{\smallskip\noindent}
\newcommand{\mn}{\medskip\noindent}
\newcommand{\bn}{\bigskip\noindent}
\setlength{\baselineskip}{17pt}

\title{Euclidean signature semi-classical methods for bosonic field theories: interacting scalar fields} 
\author{Antonella Marini}
\affiliation{Department of Mathematical Sciences, \\ Yeshiva University, 2495 Amsterdam Avenue, New York, NY 10033-3201, USA \\ and \\ Department of Mathematics, \\ University of L'Aquila, Via Vetoio, 67010 L'Aquila, AQ ITALY. \\ E-mail address: marini@yu.edu}
\author{Rachel Maitra}
\affiliation{Department of Applied Mathematics, \\ Wentworth Institute of Technology, 550 Huntington Avenue, Boston, MA 02115-5998, USA. \\ E-mail address: maitrar@wit.edu}
\author{Vincent Moncrief}
\affiliation{Department of Physics and Department of Mathematics, \\ Yale University, 217 Prospect Street, New Haven, CT  06511-8499 USA. \\ E-mail address: vincent.moncrief@yale.edu}

%EndAName

\date{January 1, 2016}  

\begin{abstract}
Elegant `microlocal' methods have long since been extensively developed for the analysis of conventional Schr\"{o}dinger eigenvalue problems. For technical reasons though these methods have not heretofore been applicable to quantum field theories. In this article however we initiate a `Euclidean signature semi-classical' program to extend the scope of these analytical techniques to encompass the study of self-interacting scalar fields in \(1+1\), \(2 + 1\) and \(3 + 1\) dimensions. The basic microlocal approach entails, first of all, the solution of a single, nonlinear equation of Hamilton-Jacobi type followed by the integration (for both ground and excited states) of a sequence of \textit{linear} `transport' equations along the `flow' generated by the `fundamental solution' to the aforementioned Hamilton-Jacobi equation. Using a combination of the direct method of the calculus of variations, elliptic regularity theory and the Banach space version of the implicit function theorem we establish, in a suitable function space setting, the existence, uniqueness and global regularity of this needed `fundamental solution' to the relevant, Euclidean signature Hamilton-Jacobi equation for the systems under study. Our methods are applicable to (massive) scalar fields with polynomial self-interactions of renormalizable type. They can, as we shall show elsewhere, also be applied to Yang-Mills fields in \(2 + 1\) and \(3 + 1\) dimensions.
\end{abstract}

\maketitle

\section{Introduction}
\label{S-intro} 
Though the Schr\"odinger representation was proven to exist for renormalizable scalar fields by Symanzik \cite{Symanzik1981} and though this conclusion is often assumed to hold for (non-abelian) gauge fields as well \cite{Luscher1992,  Gawedzki1982}, direct attempts to solve Schr\"odinger's equation for such systems are rather rare. One reason for this is surely the relative lack of suitable mathematical methods for such a task, especially by comparison with those available for the much more extensively developed path integral formalism for computing scattering amplitudes. A familiar, if not often used, exception to this is conventional Rayleigh/Schr\"odinger perturbation theory, suitably upgraded to a field theoretic setting through a regularization of the associated, formal Schr\"odinger operator and augmented by an appropriate renormalization of the physical parameters arising therein \cite{Hatfield1998}. On the other hand, a number of important physical effects are known to be inaccessible to purely perturbative methods which, indeed, somewhat artificially decompose an essentially nonlinear system into linear and interactive components and attempt to recover the potentially crucial effects of the nonlinearities gradually, in the form of series expansions in an associated coupling constant. By contrast, Euclidean signature semi-classical methods, as we shall discuss them here, seek to take the effects of \emph{quantization} increasingly into account, in the form of power series expansions in Planck's constant, while keeping the basic nonlinearities and invariances of an interacting system fully intact at every level of the analysis.

Though conventional semi-classical methods have a long and distinguished history of application to quantum mechanical problems, the authors argued in a recent paper that a less familiar  variant of these methods might be especially suitable for extension to field theoretic systems \cite{MMM-QM}. In particular their article showed, for certain systems of nonlinear quantum oscillators, how one could solve the basic inverted-potential-vanishing-energy Hamilton-Jacobi equation, that arises naturally therein, for a globally defined `fundamental solution' through an application of the direct method of the calculus of variations and then establish the smoothness of this solution through the use of a (Banach space) version of the implicit function theorem. Higher order quantum corrections to this fundamental solution could then be computed, for both ground and excited states, through the direct integration of a sequence of linear, first order transport equations derived from Schr\"odinger's equation and corresponding corrections to the energy eigenvalues obtained from the natural demand that the associated wave function corrections be globally smooth. 

In contrast to the more familiar Wentzel, Kramers and Brillouin (or WKB) semi-classical methods of the physics literature \cite{BrackBhaduri2008, Orozio1988}, this approach avoids the necessity to match solutions across a boundary between classically allowed and classically forbidden regions -- a serious complication for higher dimensional problems -- and, for related reasons, also avoids the necessity of making intricate Maslov-type corrections for the occurrence of caustics \cite{MaslovFedoriuk1981, Lazutkin1993} which, for the convex potential energies under study there were, in fact, non-existent \cite{MMM-QM}. 
\footnote{For more general, non-convex potentials on the other hand, for which caustics, though far less intrusive than in the conventional approach, could still occur, it is plausible that the scope of the authors' approach could be further enhanced by a suitable  incorporation of such Maslov-type techniques.}
 Though similar in spirit to certain elegant and rigorous approximation methods developed previously in the microlocal analysis literature \cite{DimassiSjostrand1999, Helffer1988} the authors' approach also avoids, unlike this earlier innovative work, the reliance on mathematical results and techniques whose applicability is apparently limited to finite dimensional problems (\cf the discussion in the `Concluding Remarks' section of Ref.~\cite{MMM-QM}).

A motivation for carrying out the analysis first in the relatively more elementary context of finite dimensional systems was the fact that one could thereby temporarily sidestep the intricate complications of regularization and renormalization while still developing the general features of the method for subsequent application to field theories. In fact the authors' approach naturally splits into a single nonlinear, but essentially `classical'  problem (the solution of the Hamilton-Jacobi equation alluded to above) and a sequence of linear calculations of quantum corrections wherein, for field theoretic problems, the issues of regularization and renormalization would only then first arise. Borrowing the language of conventional field theory one could say that solving the aforementioned Hamilton-Jacobi equation represents the `tree level' contribution to the approximation whereas subsequent integration of the relevant transport equations would correspond to evaluating successive `loop corrections' thereto. However, our techniques, even when applied to relatively elementary quantum mechanics problems, do not simply reproduce the standard results of  Rayleigh/Schr\"odinger perturbation theory (except in the special, `unperturbed' case of linear oscillators where they regenerate the well-known exact solutions) but instead yield substantially improved approximations to the actual wave functions, capturing in particular (even at the lowest orders of calculation) the more rapid-than-gaussian decay known rigorously to hold for such nonlinear oscillatory systems. Thus our application of this (`tree level/loop correction') terminology should not be interpreted as having a precise correspondence with the conventional usage. The expectation, however, that one should ultimately be able to generate much more accurate approximations to the Hilbert space of quantum states for certain quantized fields is one of the principal motivations of the present work.

The main aim of this paper is to extend the analysis of deriving a fundamental solution to the relevant Hamilton-Jacobi equation and establishing its properties to the important cases of $\Phi^4$ scalar fields in $4$-dimensional Minkowski spacetime, to $\Phi^p$ fields with exponent $p = 4$ or $6$ in $3$-dimensional Minkowski space, and with $p$ an arbitrary, positive even integer greater than $2$ in $2$-dimensional Minkowski space. For technical reasons we focus specifically on massive fields but also allow for intermediate powers in the (polynomial) potential function, provided the function is convex. With the ansatz we make for the ground state wave functional for such problems, a Euclidean signature (and vanishing energy) variant of the field theoretic Hamilton-Jacobi equation naturally emerges to determine the first (or `tree') approximation to (the logarithm of) this functional. While this Hamilton-Jacobi equation has occasionally been written down and discussed heuristically in the literature we present here a systematic, rigorous method for proving the existence and smoothness of its aforementioned, global `fundamental solution' in the cases of interest. These features are crucial to establish before one can proceed to the computation of quantum corrections.

For each of the scalar fields under study we prove the global existence and smoothness (in a suitable function space setting) of the relevant Hamilton-Jacobi functional, $S_{(0)} [\varphi]$, by exploiting the smoothness, coerciveness and convexity features satisfied by their (functional) potential energies -- direct analogues of properties assumed to hold in finite dimensions for the potential energies considered in Ref.~\cite{MMM-QM}.  We also discuss how our tree approximation for these ground state wave functionals already begins to capture the more-rapid-than-gaussian decay that should surely hold for the exact solutions. By contrast note that, to any finite order, conventional Rayleigh/Schr\"odinger theory would generate instead a (functional) polynomial multiplied by a gaussian and thus an approximate wave functional that decays more \emph {slowly} than the corresponding gaussian. Since, in our setup, the squared modulus of the ground state wave functional provides the natural integration measure for the associated Hilbert space of quantum states -- a delicate issue in any infinite dimensional setting -- it is quite encouraging that our approach exhibits this vital non-Fock-like behavior already at leading order.

A remarkable result from Ref.~\cite{MMM-QM} showed, in finite dimensions, that the first quantum `loop' correction to this tree approximation also made a natural, \emph{geometric} contribution to the Hilbert space integration measure by precisely cancelling the Jacobian determinant that arises from the transformation to so-called Sternberg coordinates for the associated (inverted potential) mechanics problem and thus leaving the Sternberg-coordinate Lesbegue measure in its place. Since Sternberg coordinates, by construction, linearize the gradient (semi-) flow generated by the corresponding fundamental solution of the (inverted-potential) Hamilton-Jacobi equation, the result is an elegant simplification of the natural integration measure for the quantum states if in fact Sternberg coordinates are employed therein. This geometric interpretation of the first quantum loop correction is reminiscent of that of the van Vleck-DeWitt-Morette determinant that arises in the conventional semi-classical approximation to the propagator for  Schr\"odinger's equation (see Ref.~\cite{Morette}, and Ch. 18 of Ref.~\cite{MoretteCartier}). It would be most interesting to determine whether this natural geometric aspect of the first quantum loop correction can be preserved for infinite dimensional problems. We only remark here that the analogue of the Sternberg transformation has already been studied for some interesting field theoretic problems in Refs.~\cite{Shatah1985, EckmannEpsteinWayne1993}.

While $\Phi^4$  fields in $4$ dimensions are often believed (though still not rigorously proven \cite{Podolsky2010, Suslov2008}) to renormalize to (trivial) free fields, such a conclusion, if true, should only emerge at the level of the higher order quantum, `loop' corrections that we do not consider here. The necessity to regularize otherwise ill-defined functional Laplacians only arises in our approach at the level of the transport equations for these higher order corrections and we sidestep such complications here by considering only the construction of the aforementioned fundamental solutions. In any case the fields we consider are, in lower dimensions, rigorously known to be non-trivial \cite{GlimmJaffeI, GlimmJaffeII, GlimmJaffeIII, GlimmJaffeIV, FeldmanOsterwalder1976} and we propose to deal with their higher order, `loop' corrections, for both ground and excited states, in subsequent work.

Our program of developing Euclidean signature semi-classical methods for bosonic field theories was originally designed with an application to (non-abelian) gauge theories as its primary aim and we are pursuing that objective in parallel with the present one \cite{MMM-YM}. Much of the mathematical technology needed for handling the associated Dirichlet problems for (Euclidean signature) Yang-Mills fields, in three and four spacetime dimensions, is already at hand in a form that is especially suited to the applications we have in mind \cite{Marini1992, Uhlenbeck1983, Uhlenbeck1982, Sedlacek1982}. Semi-classical methods, in the sense that we use the term, have the very natural feature of maintaining the nonlinearities and (nonabelian, gauge) invariances of a Yang-Mills system intact at every stage of the analysis rather than attempting to restore, gradually by perturbative expansion, those interactions and invariances that were broken at the outset by linearized approximation. They have the (closely related) advantage of dealing rigorously, even at leading order, with global features of the nonlinear, quantum dynamics rather than merely attempting to reinstate those global aspects perturbatively, through expansions about radically simplified linearized approximations. They do not, at present, have the explicit calculational power of conventional techniques but our view is that Euclidean signature semi-classical methods may well be able to shed mathematical light on those features of quantum gauge theories that are  believed to be inaccessible, in principle, to perturbative analysis.

\section{Overview and mathematical setting}
\label{S-setting}
The formal Schr\"odinger operator for a scalar field in $n + 1$ dimensions, of the type that we shall consider, is given by
\be
\label{23}
\hat H = \int_{\R^n} \left\{-\frac{\hbar^2}{2}\,\frac{\delta^2}{\delta \varphi^2( {\bf x^\prime})} + \frac{1}{2}\nabla^\prime\varphi({\bf x^\prime})\cdot\nabla^\prime\varphi({\bf x^\prime}) + \mathcal P(\varphi({\bf x^\prime})) \right\} \,d{\bf x^\prime}\,,
\ee
where ${\bf x^\prime}\in\R^n$, $\nabla^\prime$ designates the gradient on $\R^n$ and the self-interaction polynomial $\mathcal P$ is assumed to be convex, to include a mass term and to have its degree constrained by the spatial dimension under study (limited herein  to $n = 1, 2$ or $3$) as clarified further below. These assumptions are such as to 
 include the case of `massive'  $\Phi^4$ theory, for which $\mathcal P(\Phi)=1/2 \,m^2 \,\Phi^2 +\lambda \,\Phi^4$ with $\lambda, m^2>0$. The functional Laplacian is of course ill-defined, even acting on smooth functionals (since the Hessian of such a functional need not be trace class) and requires regularization for its proper definition. The influence of this regularization however, will only be felt at the level of computing the quantum `loop' corrections and not for the determination of the `fundamental solution', $S_{(0)} [\varphi(\cdot)]$, to the Euclidean signature-vanishing-energy variant of the Hamilton-Jacobi equation,
\be
\label{24}
\int_{\R^n} \left(\frac{1}{2}\frac{\delta S_{(0)}}{\delta\varphi( {\bf x^\prime}) }\,\frac{\delta S_{(0)}}{\delta\varphi( {\bf x^\prime}) } -\frac{1}{2}\nabla^\prime\varphi({\bf x^\prime})\cdot\nabla^\prime\varphi({\bf x^\prime}) - \mathcal P(\varphi({\bf x^\prime}) ) \right) \,d{\bf x^\prime}=0\,,
\ee
that arises at lowest order from substituting our ansatz
\be
\label{25}
\zero\Psi_\hbar [\varphi(\cdot)] = N_\hbar \,e^{-{S}_\hbar[\varphi(\cdot)]/\hbar}
\ee
for the ground state wave functional into the Schr\"odinger equation
\be
\label{26}
\hat H\,\zero\Psi_\hbar = \zero E_\hbar \zero\Psi_\hbar
\ee
and demanding that the latter hold order-by-order in powers of $\hbar$ relative to the formal expansions
\be
\label{S}
S_{\hbar}[\varphi(\cdot)] \simeq S_{(0)}[\varphi(\cdot)] + \hbar S_{(1)}[\varphi(\cdot)] + \frac{\hbar^{2}}{2!} S_{(2)}[\varphi(\cdot)] + \dots + \frac{\hbar^{k}}{k!} S_{(k)}[\varphi(\cdot)] + \dots 
\ee
and
\be
\label{E}
	\zero{E}_{\hbar} \simeq \hbar \left(\zero{\mathcal{E}}_{(0)} + \hbar \,\zero{\mathcal{E}}_{(1)} + \frac{\hbar^{2}}{2!} \,\zero{\mathcal{E}}_{(2)} + \dots + \frac{\hbar^{k}}{k!} \,\zero{\mathcal{E}}_{(k)} + \dots\right)\,.
\ee
In the foregoing formulas $\varphi(\cdot)$ designates a real-valued distribution on  $\R^n$ that can be interpreted (for convenience, with respect to our method of solution for $S_{(0)}[\varphi(\cdot)] $)  as boundary data prescribed, arbitrarily, on the  hypersurface 
$\{x^0\equiv ct\equiv t= 0\}$ in units for which the speed of light $c=1$ of (Euclidean)  $\R^{n+1} = \{{\bf x} \equiv(t, {\bf x}^\prime)\,:\, t\in\R\,,\;\bf x^\prime\in\R^n\}$ for a real spacetime scalar field $\Phi$ defined on the half-space  $(-\infty, 0] \times\R^n$.  These are the analogues of the boundary data {\bf x} and curves $\gamma$ defined on the half-line for the mechanics problem dealt with in Ref.~\cite{MMM-QM}.

As in the mechanics problem we seek to construct a `fundamental solution' to Eq. \eqref{24}  by first proving the existence of unique minimizers $\Phi_\varphi$ for the \emph{Euclidean signature} action functional
\be
\label{29}
\mathcal I_{es} [\Phi] \equiv \int_{\R^n} \int_{-\infty}^0\left\{ 1/2\, (\partial_t\Phi)^2  + 1/2\, \nabla^\prime\Phi\cdot\nabla^\prime\Phi + \mathcal P(\Phi)\right\} \,dt\,d{\bf x^\prime}\,,
\ee
for arbitrary boundary data $\varphi$ specified at $t = 0$, and then setting 
\[{S}_{(0)}[\varphi (\cdot)] = \mathcal I_{es} [\Phi_\varphi] \,.\]
The minimization procedure is carried out for $\Phi$ in the space of distributions 
\be
\label{space}
\ma(\varphi)\equiv\{\Phi\in H_1 (\nst)\,:\, \Phi= \varphi \mbox{ at }
\ntzero\,\},
\ee
for fixed arbitrary $\varphi\in\bb$, with 
\be\label{bspace}
 \mathcal B\equiv \{\varphi = \tr \, \tilde\varphi\,, \mbox { with } \tilde \varphi\in  H_{{3}/{2}}(\nst)\}= H_{1}(\ntzero)\,.
\ee 
Notice that the equality of spaces in \eqref{bspace}  holds because the Trace map
\[\tr\,:\,  H_{{3}/{2}}(\nst)\to H_{1}(\ntzero)\] 
 is onto; \cf for example Theorem 7.53 and Remark 7.56 in Ref.~\cite{adams}. (Generally, one loses ${1/p}$ orders of differentiability when restricting elements of $W_{k;p}(\nst)$ to the boundary of the domain, and the trace map 
 $\tr\,:\,  W_{k;p}(\nst)\to W_{k-1/p;p}(\ntzero)$ is onto for these domains. Here, because the Sobolev exponent (implicit in the use of $H_k$ spaces) is $2$,  we lose $1/2$ derivative.) 
As a side remark, notice that $\tilde \varphi$ can always be chosen to be smooth in the interior (the function $\hat \varphi$ of Lemma \ref{L-harmonic} in \S  \ref{S-ellreg} below is an example of an extension smooth in the interior). 
Notice also that, by the Trace theorem, $ H_{1}(\ntzero)$ may not seem to be the \emph{natural} choice for the space of boundary data for a minimization procedure that takes place in $H_1(\nst)$. Nonetheless, the higher degree of boundary regularity imposed is needed for the Hamilton-Jacobi equation to be well defined (\cf \S  \ref{S-conserv} below), as well as  for the arguments of the regularity theory, in order to obtain an initial improvement at the first step of the boot-strapping (\cf \S  \ref{S-ellreg} below). 

\sn The precise assumptions on the polynomial $\mathcal P(\Phi)$ are that
\be\label{poli} 
\mathcal P(\Phi)\equiv\sum_{j=2}^k a_j\Phi^j\,,\mbox{ with } k \mbox{ even}\,,\ee
and
\be\label{expcrit}   k\leq\frac{2d}{d-2}\equiv \frac{2n+2}{n-1} \;\mbox { if } \,n=2, 3\,.
\ee
Here $d= n+1$ is the dimension of the domain and no restriction from above on  $k$ is necessary if $n=1$. The number ${2d}/{(d-2)}$ is the so-called \emph{critical exponent}, as it constitutes a borderline case for the elliptic theory.  

Furthermore, we assume the convexity condition
\be\label{point-convex}
\mathcal P^{\prime\prime}(z)\equiv \sum_{j=2}^k j(j-1) a_jz^{j-2}\geq 0\,,\qquad \forall z\in\R\,,
\ee
and the condition 
\be
\label{a2}
a_2>0 \,,
\ee 
in order to rule out the case of  massless $\Phi^4$ theory, as well as the analogous cases in $2+1$ and $1+1$ dimensions,  and to guarantee coreciveness of $\ie$ (\cf \S \ref{S-E!} below). 

Because 
$$H_1(\nst)\subset L^p(\nst),\mbox{ for } 2\leq p\leq \frac{2(n+1)}{n-1}\equiv \frac{2d}{d-2}\,,$$
for  $d=n+1>2$, and $\forall p\geq 2$ 
 if $d=2$, the Euclidean action functional \eqref{29} is well-defined. 

For convenience, we will often rewrite \eqref{29} as 
\be \label{JE}
\ie [\Phi] \equiv \int_\nst
\left(\frac{\vert\nabla\Phi\vert^2}{2} + \mathcal P(\Phi)
\right) d{\bf x}\,,\ee
in which $d{\bf x}\equiv dx^0\,d{\bf x^\prime}$ is the volume element in $\nst$, ${\bf x}\equiv (x^0, {\bf x^\prime})\equiv
(x^0, \dots x^n)\in \nst$,  and 
$\nabla$ is the gradient in $n+1$ dimensions.

In the following, we will also denote simply by ${S}$ the functional ${S}_{(0)}$, our candidate fundamental solution to the Euclidean-signature-vanishing-energy variant of the Hamilton-Jacobi equation, 
 that is,  
\ba\label{S}
{S} \equiv {S}_{(0)}\,:\,&  \mathcal B\to \R \notag\\
& \varphi\mapsto \inf_{\Phi\in \ma(\varphi)}\ie[\Phi]= \ie[\Phi_\varphi] \,.
\end{align}
Although existence and uniqueness for the 
absolute minimizer $\Phi_\varphi$ of $\ie$ in the space $\ma(\varphi)$
are not  necessary conditions for the definition of ${S}$, non existence would be  (obviously) an obstruction to its  continuity and non uniqueness would be an obstruction to its differentiability. The existence and uniqueness of such a minimizer is proved in the next section.

Then, a candidate for the \emph{tree approximation} to the ground state wave functional for the $\mathcal P (\Phi)$ theory just described is the functional
\be\label{Omega}
\Omega_0(\varphi) \equiv \mathcal N e^{- {S}[\varphi]/ \hbar}\,,
\ee
where $\mathcal N$ is a normalization constant.

\section{Existence and uniqueness of a minimizer for $\ie$}
\label{S-E!}
Existence of a minimizer for the functional $\ie$ (defined in \eqref{29} or \eqref{JE}) on $\ma(\varphi)$ will follow if we can prove that such functional  is coercive and weakly lower (sequentially) semicontinuous  (\cf Theorem 1.2.5 of Ref.~\cite{Blanchard1992}), while
uniqueness of the minimizer will follow if we can show that $\ie$, which is globally defined on the convex space \(\mathcal A (\varphi)\), is strictly convex, that is, 
for any \(\Phi_1, \Phi_2 \in \mathcal A (\varphi)\), \(\Phi_1 \neq \Phi_2\), \(0 < \lambda < 1\) one has 
\begin{equation}\label{sconvex}
\ie [\lambda \Phi_1 + (1 - \lambda) \Phi_2] < \lambda\ie[\Phi_1] + (1 - \lambda) \ie[\Phi_2]
\end{equation}
(\cf Theorem 1.1.3 of Ref.~\cite{Blanchard1992}).
\subsection{Properties of the functional $\ie[\Phi]$}
\label{SS-properties}
{\em Coerciveness.} An estimate from above of the absolute value of the functional $\ie$ in terms of the $H_1$ norm of $\Phi$, such as 
\ba
\label{<}
&\vert\ie[\Phi]\vert \leq \int_{\nst } \left( \frac{\vert\nabla\Phi\vert^2}{2} + \sum_{j=2}^k\vert a_j\vert\,\vert\Phi\vert^j
\right)  dx^0\,d{\bf x^\prime} \notag\\
\qquad&\,\leq \sum_{j=2}^k\, C\Vert \Phi\Vert^j_{H_1(\nst )}\,\leq \,C_1\, \max\,\{\Vert \Phi\Vert^2_{H_1(\nst)}, \Vert \Phi\Vert^k_{H_1(\nst)}\}\,,\notag\\
\end{align}
is always satisfied, 
as it follows from the embedding $H_1(\nst )\subset L^p (\nst )$, for $2\leq p\leq k\leq 2d/(d-2)$, and does not require any hypotheses on the coefficients $a_j$. 
A reverse type of inequality, namely, an estimate from below of the absolute value of the functional $\ie[\Phi]$ in terms of the $H_1$-norm of $\Phi\in\ma(\varphi)$ such as \be\label{coerciveness}    %here
\Vert \Phi_\varphi\Vert^2_{H_1(\nst)}\leq C_2\, \ie[\Phi_\varphi]\equiv C_2\int_\nst
\left(\frac{\vert\nabla\Phi_\varphi\vert^2}{2} + \sum_{j=2}^k a_j{\Phi_\varphi}^j
\right) dx^0\,d{\bf x^\prime}\,,
\ee
cannot be obtained for these polynomial theories whenever the lowest order coefficient  $a_2$ vanishes. This is the case, in particular, for the massless $\Phi^4$ theory.
For the massive $\Phi^4$ theory instead, and, in general, for the polynomial theories with positive even coefficients and vanishing odd coefficients,  \eqref{coerciveness} can be obtained immediately.  In the case of $\Phi^4$ theory with strictly positive mass $(a_2\equiv m^2/2>0$, $a_3=0$, $a_4\equiv\lambda \geq 0)$, for example, it suffices to take $C_2 =\max\{2, 2 (m^*)^2/m^2\}$, if we define $\Vert f\Vert^2_{H_1(\st)}\equiv \int_\st \left(\vert\nabla f\vert^2+ (m^*)^2 f^2\right)\,d\bf x$ (notice that $m$, $m^*$ have dimension 1/length). Property \eqref{coerciveness} is commonly referred to as \emph{coerciveness} of the functional $\ie$.

A more general condition, sufficient to guarantee coerciveness of $\ie$, which allows for the presence of intermediate coefficients (of any sign), is given by the inequality
\be\label{coerciv.cond.}\mathcal P(\Phi) \geq C\,\Phi^2\,,\qquad C>0\,.
\ee
A simple explicit example of a coercive $\ie[\Phi]$ to illustrate this condition is obtained by specifying the polynomial term to be
$$\mathcal P(\Phi) = (\alpha\Phi +1)^2\Phi^2 + C\Phi^2\,,$$
in which  $C$ is taken to be positive and $\alpha\neq 0$.

We observe that condition  \eqref{coerciv.cond.} is guaranteed by the assumptions \eqref{point-convex} and \eqref{a2}. In fact, such assumptions guarantee that $\mathcal P (z) =  \mathcal G(z)\, z^2$ with 
\be\label{coerciv.cond2}
\mathcal G(z)\equiv \sum_{j=2}^k a_j z^{j-2}>0 \quad\forall z\in\R\,.
\ee
To see this, observe that $\mathcal G(0) = a_2>0$ (by \eqref{a2}) and that,  in order to rule out a change in the convexity of the  polynomial $\mathcal P$, no additional real roots may exist besides $z=0$, which is a minimum for $\mathcal P$;  thus, \eqref{coerciv.cond2} holds. 
Positive definiteness of $\mathcal G$  further implies that it be bounded away from zero, as it must achieve a positive absolute minimum in $\R$. In $4$ dimensions for example, $\mathcal G (z) \geq (-a_3^2+4a_2 a_4)/4a_4>0$,  yielding  $\mathcal P(\Phi ({
\bf x}))\geq (-a_3^2+4a_2 a_4)/4a_4 \, \Phi^2 ({
\bf x})$,  $\forall {\bf x}\in \st $.

\bn{\em Weak-lower (sequential) semicontinuity.}
A sufficient condition for the weak lower semicontinuity of the functional $\ie$  is the (non-strict) positivity of its second Frech\`et derivative (\cf for example Lemma 2.5.1 of Ref.~\cite{Blanchard1992}), that is
\be\label{second F}
D^2 \ie[\Phi](\omega,\omega)\equiv\int_\nst\left(\nabla\omega\cdot \nabla\omega+ \mathcal P^{\prime\prime}(\Phi)\, \omega\cdot\omega\right)\, dx^0\,d{\bf x^\prime}\geq0\,,\;\forall\omega\in \ma(0)\,,
\ee
guaranteed by the condition \eqref{point-convex}.

\mn{\em Strict convexity.} Theorem 2.6.1 of Ref.~\cite{Blanchard1992} yields that, for a continuously Fr\'{e}chet differentiable functional (such as \(\ie\,: \,\mathcal A (\varphi) \rightarrow \mathbb{R}\)), the strict convexity condition \eqref{sconvex} is  equivalent to strict monotonicity of the Fr\'{e}chet derivative which, for our problem, corresponds to the following (strict) inequality,
\begin{equation}\label{eq:324}
D\ie[\Phi_1](\Phi_1-\Phi_2) - D\ie[\Phi_2] (\Phi_1-\Phi_2) > 0\,,
\end{equation}
holding for all \(\Phi_1, \Phi_2 \in \mathcal A(\varphi)\) whenever \(\Phi_1\neq \Phi_2\). Written out explicitly this is equivalent to the requirement that
\ba\label{eq:325}
&\int_\nst \bigl(\nabla\Phi_1\cdot \nabla (\Phi_1 -\Phi_2) + \mathcal P^\prime(\Phi_1)(\Phi_1-\Phi_2)\bigr)\,  dx^0\,d{\bf x^\prime}+\notag\\
&-\int_\nst \bigl(\nabla\Phi_2 \cdot\nabla (\Phi_1 -\Phi_2) + \mathcal P^\prime(\Phi_2)(\Phi_1-\Phi_2)\bigr)\,  dx^0\,d{\bf x^\prime} \notag\\
&=\int_\nst \bigl(\vert\nabla (\Phi_1-\Phi_2)\vert^2+ (\mathcal P^\prime(\Phi_1)-\mathcal P^\prime(\Phi_2))(\Phi_1-\Phi_2)\bigr)  dx^0\,d{\bf x^\prime} >0,\notag\\
\end{align}
whenever \(\Phi_1-\Phi_2 \neq 0\). Since elements of \(\mathcal A (\varphi)\) assume the same values on $\ntzero$, the first integral is strictly positive for \(\Phi_1 \neq \Phi_2\) and thus a sufficient condition for the strict convexity of $\ie$ is that
\begin{equation}\label{eq:326}
\bigl(\mathcal P^\prime(\Phi_1)-\mathcal P^\prime(\Phi_2)\bigr)(\Phi_1-\Phi_2)\geq 0
\end{equation}
hold for arbitrary \(\Phi_1, \Phi_2\). A clever way to factor out $(\Phi_1-\Phi_2)$ from the first factor in \eqref{eq:326} is obtained by noting that
\begin{align}\label{eq:327}
&\mathcal P^\prime (\Phi_1) -\mathcal P^\prime(\Phi_2) = \int_{0}^{1} \left(\frac{d}{d\lambda} \mathcal P^\prime(\lambda\Phi_1+ (1-\lambda)\Phi_2)\right)\,d\lambda \notag\\
&=
(\Phi_1-\Phi_2) \int_{0}^{1} \mathcal P^{\prime\prime} (\lambda\Phi_1 +(1-\lambda)\Phi_2) \,d\lambda\,, \notag\\
\end{align}
implying that the convexity condition \eqref{point-convex} on $\mathcal P$,
sufficient for the weak lower semicontinuity of $\ie$, guarantees as well its strict convexity, yielding the uniqueness of the minimizer for arbitrarily chosen boundary value $\varphi\in\bb$.

\bn Alternatively, a direct way to proceed in order to show the uniqueness of the minimizer of $\ie$ on the space $\ma(\varphi)$ is the following. A stationary point $\Phi$ would necessarily satisfy
\ba
\label{pre-EL}
&D{\ie}[\Phi] (\omega) \equiv \lim_{t\to 0}\frac{\ie[\Phi
+ t\omega] - \ie[\Phi]}{t} \notag\\
&=\int_\nst \bigl(\nabla\Phi \cdot \nabla \omega +  \mathcal P^\prime(\Phi)\,\omega\bigr)\, dx^0\,d{\bf x^\prime}=0\,,\;\forall\omega\in\ma(0)\,.
\end{align}
By taking $\Phi=\Phi_1$, and $\Phi=\Phi_2$ in eq. \eqref{pre-EL}, where $\Phi_1, \Phi_2\in\ma(\varphi)$ are two minimizers for $\ie$, and subtracting the two formulas, one obtains
\be
\label{con-pre-EL}
\int_\nst \bigl(\nabla(\Phi_1 -\Phi_2)\cdot \nabla \omega + ( \mathcal P^\prime(\Phi_1)- \mathcal P^\prime(\Phi_2))\,\omega\bigr)\, dx^0\,d{\bf x^\prime}=0,\;\forall\omega\in\ma(0)\,.
\ee
Specifying $\omega= \Phi_1 -\Phi_2$ in \eqref{con-pre-EL}, applying the condition \eqref{point-convex} and the factorization \eqref{eq:327}, would yield strict positivity of the left hand side of eq. \eqref{con-pre-EL}, thus a contradiction, unless $\Phi_1 =\Phi_2$.

In both arguments given above to prove uniqueness, one might proceed by choosing, in place of \eqref{eq:327}, the factorization
\ba
&\mathcal P^\prime (\Phi_1) -\mathcal P^\prime(\Phi_2) =\sum_{j=2}^{k} ja_j\Phi^{j-1}_1-\sum_{j=2}^{k} ja_j\Phi^{j-1}_2=\notag\\
&= (\Phi_1-\Phi_2) \left(2a_2 + 3a_3(\Phi_1+\Phi_2) + \dots  + ka_k (\Phi^{k-2}_1 + \Phi^{k-3}_1 \Phi_2 + \dots + \Phi_2^{k-2})\right)\,.\notag\\
\end{align}
Condition \eqref{eq:326} is then 
 implied by the condition
that the hypersurface
$$
z=2a_2 + 3a_3(x+y) + 4a_4(x^2 +xy+y^2) + \dots + ka_k (x^{k-2} + x^{k-3} y +\dots y^{k-2})
$$ lie above the plane $z=0$.
In the case of the $\Phi^4$ theory (i.e. $n=3$, $k=4$, $a_2 =m^2/2 \geq 0$, $a_3=0$, $a_4=\lambda\geq 0$), eq. \eqref{con-pre-EL} becomes
\be\label{!phi4}
\int_\st \bigl(\vert \nabla (\Phi_1 -\Phi_2)\vert^2 + (m^2 + 4\lambda(\Phi^2_1 + \Phi_1 \Phi_2 + \Phi^2_2))(\Phi_1 -\Phi_2)^2 \bigr) \, dx^0\,d{\bf x^\prime}=0\,,
\ee
yielding directly $\Phi_1 -\Phi_2=0$, since the contributions of the gradient term would otherwise be strictly positive, and so would the contribution of the polynomial terms, unless $m=\lambda=0$.

\section{The Euler-Lagrange equations for the $\mathcal P(\Phi)$ theory}
\label{S-EL}
Following in their essential lines the classical references \cite{agmon, G-T}, we say that a weakly differentiable function $\Phi$ is a \emph{weak solution} to the \emph{Dirichlet Problem} ${\mathcal{D^{\,\prime}}}$,
\be
\label{wdp}\quad {\mathcal{D^{\,\prime}  }:}\qquad\left\{\begin{array}{ll}
\Delta \Phi = f \quad&\mbox{on }\nst\\
\Phi\in  \ma(\varphi)\,, 
\end{array}\right.
\ee
in which $\ma(\varphi)$ is defined in  \eqref{space}, $f$ is prescribed in $H_{-1}(\nst)$, the topological dual space of $\ma(0)$, and $\varphi$ is the \emph{weak} (or, \emph{generalized}) trace at $\ntzero$ of a function $\tilde\varphi \in H_1(\nst)$, \emph{if and only if}  
$\Phi$  satisfies 
\be\label {wdp1}
\Phi \in \ma(\varphi)\;\mbox { and } \;  \int_\nst  \nabla  \Phi \cdot \nabla \omega \,d{\bf x} + \int_\nst f \,\omega \,d{\bf x} =0\qquad \forall \omega \in \ma^\infty_c(0)\,.
\ee
Here and throughout this text, $d{\bf x}\equiv d x^0 d{\bf x^\prime}$ is the volume element in $\nst$, and 
\be\label{ainfty}
\ma^\infty_c (0) \equiv \left\{\omega\in \mathcal C^\infty_c\left((-\infty, 0]\times \R^n\right)\;:\; \omega = 0 \mbox{ at } \ntzero\right\}\,.
\ee
Notice that  the boundary value and its extension to the interior are both assumed to be $\mathcal C^\infty$  in Ref.~\cite{agmon}; but that is unnecessary. Likewise, the assumption $f\in H_{-1}$, in replacement of $f\in L^2$, will suffice.

{\bf \emph{Observations.}}   One should point out that Definition \eqref{wdp} rests on the following facts: 1) for  $\Phi$ a classical solution to \eqref{wdp}, integration against a function $\omega\in  \ma^\infty_c(0)$ followed by integration by  
parts, yields \eqref {wdp1}, as the boundary integrals -- the component over the hyperplane $\ntzero$ and the component at infinity -- both vanish under our assumptions; conversely, a smooth solution $\Phi\in \ma(\varphi)\cap\mathcal C^\infty(\cnst)$ of  \eqref {wdp1}  is also a classical solution to system \eqref {wdp}; 2) if  $\Phi\in \ma(\varphi)$ is  a limit (strong or weak) of a sequence of smooth functions $\Phi_j$ satisfying eq. \eqref {wdp1}, then also $\Phi$ satisfies eq. \eqref {wdp1} (that is, eq. \eqref {wdp1} is preserved in the limit).
Notice that  the space $\ma^\infty_c(0)$ is dense in $\ma(0)$, just as $\mathcal C^\infty_c (\R^n)$ is dense in $H_1(\R^n)$  in all dimensions $n$,  and one might have as well assumed  that  \eqref {wdp1} hold $\forall \omega \in \ma(0)$. A quick calculation would then show that, in the case $f\equiv 0$,  a solution of  \eqref {wdp1} would also be the (unique) minimizer in $\ma(\varphi)$  of $\int_\nst  \nabla\Phi\cdot\nabla\Phi\,d{\bf x}$.
Notice, also, that Definition \eqref{wdp} implicitly defines the \emph{weak Dirichlet Laplacian on $\ma(\varphi)$} through the identification $\Delta\Phi\simeq \int_\nst  -\nabla  \Phi \cdot \nabla (\cdot) \,d{\bf x}$ (that is, we regard  $\Delta\Phi$, for each $\Phi\in \ma(\varphi)$, as an element of $H_{-1}(\nst)$).

\mn
Returning to our problem, we say that a weakly differentiable function $\Phi$ is a \emph{weak solution} to the \emph{nonlinear Dirichlet Problem} $\mathcal{D^{\,\prime\prime}}$,
\be
\label{wnl}\quad \mathcal{D^{\,\prime\prime}:}\qquad\left\{\begin{array}{ll}
\Lambda  \,\Phi \equiv - \Delta \Phi +\mathcal  Q (\Phi) =0\quad&\mbox{on } \nst  \\
\Phi\in  \ma(\varphi)\,,
\end{array}\right.
\ee
in which $\mathcal Q(\Phi)$ is a (possibly) nonlinear function of $\Phi$ satisfying $\mathcal Q(\Phi)\in H_{-1}(\nst)$, and  $\varphi$ is the \emph{weak} (or, \emph{generalized}) trace at $\ntzero$ of a function $\tilde\varphi \in H_1(\nst)$, \emph{if and only if}  
$\Phi$  satisfies 
\be\label {wnl1}
\Phi \in \ma(\varphi),\; \mathcal Q(\Phi)\in H_{-1}(\nst),\mbox { and } \int_\nst  \left(\nabla  \Phi \cdot \nabla \omega  + \mathcal Q (\Phi) \,\omega \right)\,d{\bf x} =0\; \forall \omega \in \ma^\infty_c(0)
\ee
In the definition above, $\Lambda$  is a nonlinear operator with image identifiable with a subset of  $H_{-1}(\nst)$. That is, 
\ba
\label{Lambda}
\Lambda\;:\;\ma(\varphi) &\to H_{-1}(\nst)\,\notag\\
\Phi &\mapsto\Lambda \Phi \equiv - \Delta \Phi + \mathcal Q  (\Phi)\simeq  \int_\nst\bigl(\nabla\Phi\cdot\nabla (\cdot)+\mathcal Q(\Phi)(\cdot)\bigr)\,d{\bf x} \,.
\end{align}
Here, $\simeq$ represents the aforementioned  identification. 
The observations following Definition \eqref {wdp}-\eqref {wdp1} all generalize to this case, after replacing $\Delta$ by the operator $\Lambda$ and having provided hypotheses that guarantee that the  integral in  \eqref{wnl1}(and, in particular, its second summand) be finite when evaluated on any $\omega\in \ma(0)$.

\sn If $\mathcal P(\cdot)$ is a polynomial satisfying the assumptions described in Secs. \ref{S-setting}, \ref{S-E!},  one can take $\mathcal Q=\mathcal P^\prime$ in \eqref{wnl}, or in \eqref{wnl1}. In fact, in that case, the condition  $k\leq {2d}/{(d-2)}$ ($d\equiv n+1$) guarantees   
 that 
\[\left|\int_\nst \mathcal P^\prime(\Phi)\, \omega\,d x^0\, d{\mathbf x^\prime}\right|\,\leq C\, \max\,\left\{\Vert \Phi\Vert_{H_1(\nst)}, \Vert \Phi\Vert^{k-1}_{H_1(\nst)}\right\}\,\Vert\omega\Vert_{H_1(\nst)}\,,
\]
in which the constant $C$ depends on the coefficients of the polynomial $\mathcal P$ (this is seen by applying  
 H\"older inequalities and the Sobolev embedding 
 $H_1(\nst )\subset L^p (\nst )$ for $2\leq p\leq {2d}/{(d-2)}$; \cf  Ref.~\cite{adams}).  
 Thus for all $\Phi\in\ma(\varphi)$, $\omega\in\ma (0)$, one has 
\ba 
\label{Lambda1}
&\left| \Lambda\Phi(\omega)\right|= \left|\int_\nst\bigl(\nabla\Phi\cdot\nabla\omega +\mathcal P^\prime(\Phi)\,\omega\bigr)\,d {\mathbf x}\,\right|\notag\\
&\leq C\, \max\,\left\{\Vert \Phi\Vert_{H_1(\nst)}, \Vert \Phi\Vert^{k-1}_{H_1(\nst)}\right\}\,\Vert\omega\Vert_{H_1(\nst)}\,,
\end{align}
for some constant $C$ independent of $\Phi$ or $\omega$, and
\be
\label{estLambdab}
\Vert \Lambda\Phi\Vert_{H_{-1}(\nst)}\equiv\sup_{\omega\in\ma (0)}
\frac{\left| \Lambda\Phi(\omega)\right|}
{\Vert\omega\Vert_{H_1(\nst)}}\leq C\, \max\,\left\{\Vert \Phi\Vert_{H_1(\nst)}, \Vert \Phi\Vert^{k-1}_{H_1(\nst)}\right\}\,.
\ee
So,
$\Lambda\Phi$ is a continuous linear operator on $\ma(0)$ (that is, $\Lambda\Phi\in H_{-1}(\nst)$). 

\sn
In addition, even if the function $\mathcal Q$ in \eqref{wnl} is a more general continuous function of a real variable, setting $\mathcal Q(\Phi) = \mathcal P^\prime(\Phi)$, still under the assumption $\mathcal P^\prime(\Phi)\in H_{-1} (\nst)$, one can see that a  stationary point (not guaranteed to exist)  $\Phi\in\ma(\varphi)$ of $\mathcal J[\Phi]\equiv \int_\nst  \left(\nabla\Phi\cdot\nabla\Phi + \mathcal P(\Phi)\right)\,d{\bf x}$ would also be a weak solution  to the nonlinear Dirichlet problem \eqref{wnl} (that is, a solution to \eqref{wnl1}), and vice versa (by a density argument).
In fact, for $\Phi\in \ma(\varphi)$, $\omega \in\ma(0)$, one has
\ba & D\mathcal J[\Phi](\omega)\equiv\lim_{\lambda\to 0} \frac{1}{\lambda}\left\{\int_\nst  \left(\nabla(\Phi + \lambda\omega) \cdot\nabla (\Phi + \lambda\omega)+ \mathcal P(\Phi+ \lambda\omega)\right)\,d{\bf x} \right.\notag\\
&\left. -\int_\nst  \left(\nabla\Phi\cdot\nabla\Phi + \mathcal P(\Phi)\right)\,d{\bf x} \right\}=
\int_\nst\bigl(  \nabla\Phi\cdot\nabla\omega  + \mathcal P^\prime (\Phi) \,
\omega\bigr)  \,d{\bf x}\,;\notag\\
\end{align}
thus, $\Phi$ is a stationary point  of $\mathcal J[\Phi]$ \emph{if and only if} it satisfies $\int_\nst\bigl(  \nabla\Phi\cdot\nabla\omega  + \mathcal P^\prime (\Phi) \,
\omega\bigr)  \,d{\bf x}=0\;\forall \omega\in \ma(0)$, or, equivalently, it satisfies  the \emph{weak Euler-Lagrange equations}  \eqref{wnl}. 

In the case $\mathcal P(\cdot)$ is a polynomial as described in Secs. \ref{S-setting}, \ref{S-E!},  there exists a \emph{unique} weak solution to \eqref{wnl} (or, equivalently,  a \emph{unique} solution to \eqref{wnl1}), with prescribed boundary value $\varphi\in  \mathcal B$, and  such  solution is the \emph{unique}  minimizer of $\mathcal J=\ie$ (\cf the arguments following eq. \eqref{con-pre-EL}).  
 
\sn 
As already mentioned, the observation after eq. \eqref{ainfty} holds here as well, after replacing $\Delta$ by the operator $\Lambda$. So, in particular, for $\Phi$  a smooth minimizer of $\ie$,  integration by parts   gives 
\ba
\label{difference}
&0= \int_\nst\bigl( \nabla\Phi\cdot\nabla\omega  + \mathcal P^\prime (\Phi) \,
\omega\bigr)  \,d{\bf x} = \int_\nst\bigl((\Lambda \Phi)\,\omega  +
 \nabla \cdot \,\omega\nabla\Phi \bigr) \,d{\bf x} =
\notag\\
&\int_\nst (\Lambda \Phi)\,\omega \,d{\bf x} +
\int_\ntzero
\omega\nabla\Phi\cdot (1, {\bf 0}) \,d{\bf x^\prime} + \lim_{R\to\infty} \int_\rinfty \omega \nabla\Phi \cdot
\frac{{\bf x}}{\Vert  {\bf x}\Vert}\,d\sigma \notag\\
&= \int_\nst (\Lambda \Phi)\,\omega \,d{\bf x}\,,
\end{align} 
in which 
 $d\sigma$ is the surface element on
$\rinfty$, and $\Lambda$ is meant in the classical sense. Here,  the vanishing of the first boundary integral in \eqref{difference} comes from having prescribed $\Phi_{|\ntzero}= (\Phi +\lambda\omega)_{|\ntzero} = \varphi$, (that is,  from having prescribed the vanishing of the variation $\omega$ at $\ntzero$); the second boundary integral (over the component at infinity) vanishes automatically because of the assumption that $\Phi,\,
\omega$ be in  $H_1( \nst)$. This follows from the density of $\ma^\infty_c (0)$ in $\ma(0)$, or can be proven directly. 

\mn
Summarizing the foregoing discussion, a minimizer $\Phi$ of $\ie$ over the space $\ma(\varphi)$ (\cf Def. \eqref{space}) satisfies the \emph{Euler-Lagrange equations}
\be
\label{EL}
\Lambda \Phi \equiv - \Delta \Phi +  \mathcal P^\prime(\Phi) =0\qquad\mbox{ on } \nst\,,
\ee
together with the boundary condition $\Phi\in\ma(\varphi)$,  
\emph{weakly}, that is, in the sense that  
\be
\label{weakEL}
D{\ie}[\Phi] (\omega)\equiv \lim_{\lambda\to 0} \frac{\ie[\Phi + \lambda\omega] - \ie[\Phi] }{\lambda}   
= \int_\nst \bigl(\nabla\Phi \cdot \nabla \omega +  \mathcal P^\prime(\Phi)\omega\bigr)dx^0d{\bf x^\prime}=0\,\forall \omega\in \ma^\infty_c(0)
\ee
For $\Phi\in\ma(\varphi)\cap\mathcal C^\infty((-\infty, 0]\times \R^n)$  this  is equivalent to being 
a \emph{classical} solution to eq. \eqref{EL}. 
Further, the unique stationary point (a minimizer) $\Phi$ for $\ie$ in $\ma(\varphi)$  is the unique solution to the nonlinear Dirichlet problem \eqref{wnl} with $\mathcal Q =\mathcal P^\prime$.

\section{Elliptic theory}
\label{S-ellreg}
Interior regularity of a solution $\Phi$ to the nonlinear Dirichlet Problem 
\be
\label{wnlP}\left\{\begin{array}{ll}
\Lambda  \,\Phi \equiv - \Delta \Phi +\mathcal  P^\prime (\Phi) =0\quad&\mbox{on } \nst  \\
\Phi\in  \ma(\varphi)\,,
\end{array}\right.
\ee
with $\mathcal P$ a polynomial satisfying the assumptions prescribed in Secs. \ref{S-setting}, and  $\varphi\in  \mathcal B$ (defined in \eqref{bspace}), 
can be obtained by means of standard bootstrapping  and is a straight-forward procedure in the case in which $\mathcal P (\Phi)$ has degree strictly less than the critical exponent (\cf \eqref{expcrit}). This is always the case in $d=n+1=2$ dimensions. For the case in which the degree of $\mathcal P (\Phi)$ equals the critical exponent in the given dimension, we need an additional initial step and a more complex procedure.  More precisely, we first exploit the existence theory to replace system \eqref {wnlP} by a linear system (freezing part of the equation), then apply a regularity lifting method by means of a contracting operator,  to obtain the improvement at the first step of the bootstrapping. Once the first improvement is obtained, we proceed as one would in the subcritical case. 
The additional difficulties pertaining to the boundary case are overcome by reflecting system \eqref  {wnlP}   across the boundary, thus transforming the problem into an interior problem.  
In order to do so, first we extend the boundary value $\varphi$ to either a function $\hat\varphi$ defined in the interior which satisfies Laplace's equation, or to a function $\Phi_L$ which satisfies the linearized problem; then we use these extensions to replace \eqref  {wnlP}  by a system with homogeneous boundary value. Additional complications arise in establishing global control of the solution over the unbounded domain $\nst$.

We prove the following theorem. 
\bt\label{T-reg1} 
Let $\Phi$ be a solution to \eqref  {wnlP} with prescribed boundary value $\varphi\in\bb$, in which we assume 
$d\equiv n+1=2$, $3$ or $4$.
Let $\hat\varphi\in H_{1}(\nst)$ be the extension of $\varphi$ satisfying Laplace's equation $\Delta \hat\varphi =0$ on $\nst$, $\Phi_L$ be the extension  of $\varphi$ satisfying the linearized problem \eqref{l} below. 
Then  $\hat\varphi, \Phi_L, \Phi\in H_{{3}/{2}}(\nst)\cap \mathcal C^\infty (\nst)$ (that is, these functions are smooth in the interior and all satisfy the best possible global estimate on $\nst$).  
 If, in addition, the boundary value $\varphi\in \mathcal C^\infty(\ntzero)$, then $\Phi\in H_{{3}/{2}}(\nst)\cap\mathcal C^\infty\left((-\infty, 0]\times \R^n\right)$, that is, $\Phi$ is smooth all the way up to and including the boundary. 
\et
In the course of the proof, we shall also see that, for general boundary data in $\varphi\in\bb$, in  $3$ dimensions the function  $\tilde\Phi\equiv\Phi - \hat \varphi$ satisfies $\tilde\Phi\in\mathcal C^1( (-\infty, 0]\times \R^2 )$, while in $4$ dimensions $\tilde\Phi\in\mathcal C^0( (-\infty, 0]\times \R^3 )$.

\sn\emph {Proof of Theorem \ref{T-reg1}:} For convenience we recall that the Sobolev Embedding Theorem on bounded domains $\Omega$ of dimension $d$ affirms  the embeddings 
\be\label{SET} L^q_h (\Omega) \subset  L^p_l(\Omega)\,,\;\mbox { 
if and only if  } h\geq l\geq 0 \mbox{  and } 
\frac{h-l}{d} +\frac{1}{p} -\frac{1}{q} \geq 0\,;\quad p,q\geq 1\,.
\ee   
Only some of these embeddings hold on unbounded domains. These comprise for example the inclusions $ L^q_h (\Omega) \subset  L^p(\Omega)$ in the additional hypothesis $q\leq p$, for $\Omega$ a domain having the cone property, in  dimension $d> hq$; \cf Theorems 7.57, 7.58 in Ref.~\cite{adams} (better regularity results hold if $d\leq  hq$).  Among these, we have repeatedly used in particular  the embeddings 
\be\label{used}
H_1(\nst)\subset L^p(\nst)\;\forall 2\leq p\leq 2d/ (d-2)\mbox{  if } d\geq 3\,, \mbox{ and }\;\forall p\geq 2   
\mbox{  if } d=2\,.
\ee
\emph{Interior regularity.} 
As already mentioned, smoothness of $\Phi$ at interior points is easier to achieve in the subcritical case. So we focus on the case in which the degree of the polynomial term $\mathcal P(\Phi)$ equals  the critical Sobolev exponent; that is,  
\be\label{crit}
k\equiv\deg \mathcal P(\Phi)= \frac{2d}{d-2}\equiv\frac{2(n+1)}{n-1}\,,\quad n=2, 3\,.
\ee
In that case, $\mathcal P^\prime(\Phi)$ is only in $L^{{2d}/{(d+2)}}_{loc}(\nst)$, yielding 
$\Phi\in L^{{2d}/{(d+2)}}_{2;loc}(\nst)\subset H_{1;{loc}}(\nst)$. (Incidentally, since the inclusion  $L^{{2d}/{(d+2)}}_{2}(\Omega)\subset H_1(\Omega)$,  for $\Omega$ a bounded domain, does not depend on the size of $\Omega$ and $\Phi$ is assumed to be in $H_1(\nst)$, this implies that $\Phi\in L^{{2d}/{(d+2)}}_{2}(\nst)$ and  $\mathcal P^\prime(\Phi)\in L^{{2d}/{(d+2)}}(\nst)$.)  Unfortunately,  the latter is a borderline inclusion, yielding no improvement at the first step of the bootstrapping. To overcome this problem, one utilizes the existence theory to  freeze part of the coefficients by  factoring out $\Phi$ from the polynomial $\mathcal P^\prime (\Phi)$, thus rewriting the equation in \eqref  {wnlP}  as 
\be
\label{freeze}
\Lambda_\Phi \Phi \equiv -\Delta \Phi + g (\Phi)\,\Phi =0\,,
\ee
in which $\Lambda_\Phi \equiv -\Delta  + g (\Phi)\cdot$ is a linear operator. The coefficient $g (\Phi)\equiv \mathcal P^\prime (\Phi)/\Phi$ satisfies $g (\Phi)\in L^{d/2}_{loc}(\nst)$ (utilizing the embedding 
$H_1(\nst)\subset L^{2d/(d-2)}(\nst)$  and that, for $\Omega$ a bounded domain,  holds the inclusion $L^p(\Omega)\subset L^q(\Omega)$, if $p\geq q$; note that the highest order term in $g(\Phi)$ has exponent $2d/(d-2) -2 = 4/(d-2)$ ). 

The solution  $\Phi\in H_1(\nst)$ to \eqref  {wnlP}  then satisfies by construction the problem 
\be
\label{lin}\quad\qquad\left\{\begin{array}{ll}
\Lambda_\Phi u\equiv -\Delta u + g (\Phi)\,u =0
&\mbox{on }\; \nst\\
u\in\ma(\varphi)\,,
\end{array}\right.
\ee
with prescribed $\varphi\in\bb$ and 
coefficient $g (\Phi)\in L^{{d}/{2}}_{loc}(\nst)$. 

Let  ${\bf a}$ be a point in the interior of $\nst$,  $R$ and $R_1$ be positive numbers chosen so that  $0<R_1<R$ and the ball of center ${\bf a}$ and radius $R$ be contained in $\nst$; that is, 
$$B_{{\bf a};R}\equiv \{ {\bf x}\,:\, \Vert {\bf x} - {\bf a}\Vert < R\}\subset \nst\,;$$
let $\alpha_1$ be a smooth cut-off function, compactly supported in $B_{R}$, with value $1$ on $B_{R_1}$.  
Then 
$$
{\Lambda}_\Phi (\alpha_1\Phi) \equiv -\Delta (\alpha_1\Phi) + g (\Phi)\alpha_1\Phi = \alpha_1{\Lambda}_\Phi (\Phi) + \mathcal L_{\alpha_1} (\Phi) =  \mathcal L_{\alpha_1} (\Phi)\equiv f_1\mbox{ on } \nst,
$$
in which  $\mathcal L_{\alpha_1} (\Phi)= -2\,\nabla \Phi \cdot \nabla \alpha_1 - \Phi\,\Delta \alpha_1\in L^2(\nst)$. 
 Thus $\alpha_1\Phi$ satisfies 
\be
\label{cutoff1}\quad\qquad\left\{\begin{array}{ll}
{\Lambda}_\Phi  u \equiv -\Delta u + g (\Phi) u  = f_1&\mbox{on } \nst\\
\quad \quad u\in H_{1; 0} (B_{{\bf a};R})\,,
\end{array}\right.
\ee
with $g(\Phi)\in  L^{{d}/{2}}(B_{{\bf a};R})$, $f_1\in L^2(B_{{\bf a};R})$.  Using $L^2(B_{{\bf a};R})\subset L^{{d}/{2}}(B_{{\bf a};R})$  in $d=3$ or $4$  dimensions, as in the cases considered, the \emph {regularity lifting} Theorems 3.3.1 and 3.3.2 of Ref.~\cite{C} can then be applied directly, with no adaptations needed, yielding $\alpha_1\Phi$  in $L^p(B_{{\bf a};R})$  for all $p>1$. 
Thus $\Phi\in L^p(B_{{\bf a};R_1})$   for all $p>1$. 

Let $\alpha_2$ be a new smooth cut-off function compactly supported in $B_{{\bf a};R_1}$, with value $1$ on the ball 
$B_{{\bf a};R_2}$, for $0<R_2<R_1$. 
It follows from \eqref  {wnlP}   that $\alpha_2\Phi$ satisfies 
\be
\label{sys-int}\quad\qquad\left\{\begin{array}{ll}
-\Delta u = f_2\equiv-\alpha_2 \,\mathcal P^\prime(\Phi)+\mathcal L_{\alpha_2} (\Phi)\quad&\mbox{on } \nst\\
\quad \quad u\in H_{1; 0} (B_{{\bf a};R_1})\,,
\end{array}\right.
\ee
in which  $f_2\in L^2(\nst)$ (using $\mathcal P^\prime(\Phi)\in L^p(B_{{\bf a};R_1})\,\forall p>1$ and $\mathcal L_{\alpha_2} (\Phi)\in L^2(\nst))$. 
Standard elliptic theory then yields $\Phi\in \mathcal C^\infty(B_{{\bf a};R})$, for some $0<R<R_2$;  
see for example Ref.~\cite{G-T}). Because the point ${\bf a}\in\nst$ is arbitrarily chosen, this yields $\Phi\in \mathcal C^\infty(\nst)$.

\sn\emph{Regularity up to and including the boundary. First improvement.}  Once again, we focus on the critical case (\cf \eqref{crit}).  
We first prove the following lemma.
\begin{lemma} \label{L-harmonic}
Let $\hat\varphi$ satisfy 
\be
\label{Dirichlet}\quad\qquad\left\{\begin{array}{ll}
\Delta u  =  0 \quad&\mbox{ on } \;\nst\\
u\in\ma(\varphi)\,,
\end{array}\right.
\ee
with prescribed $\varphi\in\bb$. Then $\hat\varphi\in H_{{3}/{2}}(\nst)\cap \mathcal C^\infty(\nst)$. If, in addition, the boundary value $\varphi$ is smoothly prescribed, then $\hat\varphi\in\mathcal C^\infty(\inst)$. 
\end{lemma}
\emph{Proof of Lemma \ref{L-harmonic}:}  
The (unique) solution $\hat \varphi$ to \eqref{Dirichlet} is also the 
 unique minimizer of the Dirichlet integral
$
\int _\nst \vert\nabla u \vert^2 \,d{\bf x}
$
over the affine space $\mathcal A (\varphi)$, with prescribed $\varphi
\in\bb$, and its  $\mathcal C^\infty$-regularity in the interior can be found in standard references for elliptic theory; see for example Ref.~\cite{G-T}. Here we will 
 prove the global estimate $\hat\varphi\in H_{{3}/{2}}(\nst)$, as well as $\mathcal C^\infty$-regularity up to and including the boundary in the case of smooth boundary data $\varphi$. 

Let $\tilde\varphi$ be any extension to the interior of $\varphi\in\bb$, living in $H_{{3}/{2}}(\nst)$ (this exists because the trace map $H_{{3}/{2}}(\nst)\to H_{1}(\ntzero)$ is onto). The function $u\equiv \hat \varphi -\tilde\varphi$ then satisfies
\be\notag
\quad\qquad\left\{\begin{array}{ll}
\Delta u  =  f\equiv -\Delta  \tilde\varphi\in H_{-1/2}(\nst)\quad&\\
u \in\ma(0)\,.
\end{array}\right.
\ee
Its odd extension $\check u$ across $\ntzero$ satisfies weakly 
\be\label{dDirichlet}
\Delta \check u = \check f\in H_{-1/2}(\R^{n+1})\,,
\ee
in which $\check f$ is the odd extension of $f\equiv -\Delta  \tilde\varphi$ across $\ntzero$.
Equation \eqref{dDirichlet} is satisfied because odd (or even) extensions across $\ntzero$ preserve membership in $H_{-1/2}$ 
and, furthermore,  the second normal derivative of $\check u$  presents only a jump discontinuity at $\ntzero$ (no $\delta$ function is created).  
Passing to Fourier transforms one can see that equation \eqref{dDirichlet} implies that $\check u\in H_{3/2}(\R^{n+1})$, thus $u\in H_{3/2}(\nst)$. Therefore, $\hat \varphi = u+ \tilde\varphi\in H_{3/2}(\nst)$. 

In the case in which $\varphi\in\mathcal C^\infty(\ntzero)$,  the extension $\tilde \varphi$ can be chosen to be smooth, thus yielding $\check f\in L^p_{loc}(\nst)$ for all $p$. Then, eq. \eqref{dDirichlet} implies $\check u\in L^p_{2;loc}(\nst)$. At this point one can start taking tangential derivatives  of $\hat\varphi$ along $\ntzero$ and prove regularity of those. Normal derivatives are then related to tangential derivatives via Laplace's equation, thus yielding $\hat \varphi\in \mathcal C^\infty (\nst)$ for smooth boundary data $\varphi$.

This concludes the proof of Lemma \ref{L-harmonic}.
\qed

In order to continue the proof of Theorem \ref{T-reg1} we observe that 
the function $\tilde\Phi\equiv\Phi-\hat\varphi$ satisfies the problem 
\be
\label{lb}\quad\qquad\left\{\begin{array}{ll}
\Lambda_\Phi u\equiv -\Delta u + g(\Phi) u=  - g(\Phi)\hat \varphi\equiv r&\mbox{ on } \;\nst\\
u \in\ma(0)\,,
\end{array}\right.
\ee
with vanishing Dirichlet boundary conditions. 
As previously estimated, $g(\Phi)\in L^{{d}/{2}}_{loc}(\nst)$ in dimension $3$ and $4$, while $\hat\varphi\in  H_{{3}/{2}}(\nst)\subset L^{p} (\nst)$ for all $2\leq p\leq {2d}/{d-3}$ in  $d\equiv n+1 >3$ dimensions, and for all $p\geq 2$ if $d=3$. 
H\"older's inequality thus gives $r\in L^{{2d}/{(d+1)}}_{loc} (\nst)$ if $d>3$, whereas for the three-dimensional case $r\in L^p_{loc}(\threest)$ $\forall  p< 3/2$. 
Note that, if the boundary value $\varphi$ is, in addition, smooth, one has $\hat \varphi\in L^\infty_{loc}(\nst)$, thus  $r\in  L^{{d}/{2}}_{loc}(\nst)$ in $d\geq 3$ dimensions. 
 
The first equation in system \eqref{lb} can be extended by reflection across the boundary yielding  the equation  
\be
\label{doubl1}
{\tilde \Lambda}_\Phi u\equiv -\Delta u + \overline{g (\Phi)}\,u = \check {r} \quad\mbox{on } \R^{n+1}\,,
\ee
in which $\overline {g(\Phi)}$ is the even extension across the boundary of $g (\Phi)$ and  $\check r$ is the odd extension of $r$.
Because  even and odd extensions across the boundary preserve membership in $L^p$,  
$\overline {g(\Phi)}\in L^{{d}/{2}}_{loc}(\R^{n+1})$,  and $\check r\in 
L^{{2d}/{(d+1)}}_{loc}(\R^{n+1})$ if $d>3$,   $\check r\in L^{p}(\R^{3})$ $\forall p<3/2$ if $d=3$.

We define $\check \Phi$ as the odd extension across the boundary of $\tilde \Phi$. Since $\tilde\Phi=0$ on $\ntzero$,  $\check\Phi$ is a weak solution in $H_1 (\R^{n+1})$ to eq.  \eqref{doubl1}.

Let ${\bf a}=(0, a^1, a^2, a^3)$ be a fixed boundary point, $R$ any positive number,  and define the sets 
$$\mathcal U_R\equiv\{ {\bf x}=(x^0, {\bf x^\prime} )\in (-\infty, 0]\times \R^n\;:\;\Vert {\bf x} - {\bf a}\Vert < R\}\,,$$
$$B_{{\bf a};R}\equiv \{ {\bf x}\in\R^{n+1}\,:\, \Vert {\bf x} - {\bf a}\Vert < R\}\,.$$
Fix $0<R_1<R$ and let $\alpha_1$ be a smooth cut-off function, compactly supported in $B_{{\bf a};R}$, with value $1$ on $B_{{\bf a};R_1}$.  
Then,
\be
\label{cutoff}
{\tilde\Lambda}_\Phi (\alpha_1\check\Phi) = \alpha_1 {\tilde\Lambda}_\Phi (\check\Phi) + \mathcal L_{\alpha_1} (\check\Phi) = \alpha_1 \check r + \mathcal L_{\alpha_1} (\check\Phi)\equiv h_1\,,
\ee
in which $\mathcal L_{\alpha_1} (\cdot)$ is the lower order linear differential operator with smooth coefficients dependent on 
$\alpha_1$ encountered earlier. In all dimensions $d$, one has 
$\mathcal L_{\alpha_1} (\check\Phi)\in L^2(B_{{\bf a};R})\subset  L^{{2d}/({d+1})}(B_{{\bf a};R})$. Thus,  $h_1\in L^{{2d}/({d+1})} (\R^{n+1})$ in $d\geq 4$ dimensions (using also that $\alpha_1$ is supported in $B_{{\bf a};R}$) and, in particular,  
$h_1\in L^{{8}/{5}} (\R^{4})$ if $d=4$.
A similar analysis yields  $h_1\in L^{p}(\R^{3})$ $\forall p<3/2$ in $d=3$ dimensions. In conclusion, 
$\alpha_1\check\Phi$ solves the linear system
\be
\label{lin3}\quad\qquad\left\{\begin{array}{ll}
{\tilde \Lambda}_\Phi u \equiv-\Delta u + \overline{g(\Phi)} u = h_1 &\mbox{ on } \R^{n+1}\\
 \quad u\in H_{1; 0} (B_{{\bf a};R})\,.
\end{array}\right.
\ee
with  coefficients $\overline{g(\Phi)}\in L^{{d}/{2}}_{loc}(\R^{n+1})$,  $ h_1\in L^{{2d}/({d+1})} (\R^{n+1})$ in $d=n+1>3$ dimensions, $h_1\in L^{p}(\R^{3})$ $\forall p<3/2$ in $d=n+1=3$ dimensions.

Because  $h_1$ does not satisfy the hypothesis formulated in Theorem 3.3.2 of Ref.~\cite{C}, we outline and extend here the procedure carried out in that reference.
One starts by considering the equation 
\be
\label{C-1}
-\Delta u = a(x) u + b(x)\,,\; u\in H_{1; 0}(\Omega)\,,
\ee 
for $\Omega$ a bounded domain in $d>2$ dimensions.
We assume $a(x)\in L^{\frac{d}{2}}(\Omega)$ (as in Ref.~\cite{C}), but work with the weaker hypothesis $b(x)\in L^{\frac{2d}{d+1}} (\Omega)$ if $d>3$, or $b(x)\in L^{p} (\Omega)$ $\forall p< 3/2$ if $d=3$ (the corresponding hypothesis in Ref.~\cite{C} is $b(x)\in L^{\frac{d}{2}}(\Omega)$, $d>2$). For us, $\Omega \equiv B_{{\bf a};R}$. 
We prove the following Lemma.
\begin{lemma} \label{L-RL} Regularity Lifting Theorem. 
Let $u$ satisfy eq. \eqref{C-1}, with coefficients $a(x)\in L^{\frac{d}{2}}(\Omega)$ and  $b(x)\in L^{\frac{2d}{d+1}} (\Omega)$ if $d>3$, $b(x)\in L^{p} (\Omega)$ $\forall p< 3/2$ if $d=3$.  Then $u\in L^{{2d}/{d-3}}(\Omega)$ in $d>3$ dimensions $u\in L^p(\Omega)$ $\forall p\in (1,\infty)$ in $3$ dimensions.
\end{lemma}
\emph{Proof:}
Let $$\Omega_A\equiv \{{\bf x}\in \Omega \;:\;\vert a(x)\vert\geq A\}\,,$$
for $A$ a positive fixed constant. 
 We rewrite 
\be\label{aAB} a(x) = a_A(x) + a_B(x)\,,\quad \mbox{ with }
a_A(x)\equiv a(x)\mathcal X_{A}(x)\,,\quad 
a_B(x)\equiv a(x)\mathcal (1-\mathcal X_{A}(x))
\,,\ee
in which $\mathcal X_{A}$ is the characteristic function of the set $\Omega_A$.  
By applying the solution operator $T\equiv (-\Delta)^{-1}$ to both sides of \eqref{C-1}, one obtains the equation 
\be
\label{u}
u= T_A(u) + g\,,
\ee
in which $T_A(\cdot)\equiv T(a_A \,\cdot)$, $g = g_1 + g_2$, $g_1 =T(a_B \,u)$, $g_2=T(b)$. 

If $A$ is  a sufficiently large constant,  the operator 
$$T_A\,:\, L^p(\Omega) \to L^p(\Omega)$$ is a contracting operator $\forall p\in (\frac{d}{d-2}, \infty)$.
In fact, 
\be\label{contractp}
\Vert T_A (v)\Vert_p=\Vert T(a_A v)\Vert_p\leq C_1\Vert T( a_A\,v)\Vert_{{\frac{pd}{2p+d}}; 2}\leq C_2 \Vert a_A\,v\Vert_{\frac{pd}{2p+d}}\leq C_3 \Vert a_A\Vert_{d/2} \,\Vert v\Vert_p\leq \frac{1}{2} \Vert v\Vert_p\,,\ee 
in which we have used the Sobolev embedding $L^{\frac{pd}{2p+d}}_2(\Omega)\subset L^p(\Omega)$ for the first inequality; the second inequality follows from the fact that 
$$T\,:\, L^q(\Omega)\to L^q_{2;0}(\Omega)\,;\quad q>1$$ 
is a bounded operator (the requirement $p> d/(d-2)$ entails $q=dp/(2p+d)>1$);  the third inequality is H\"older's inequality;  the last inequality holds because $a\in L^{\frac{d}{2}}(\Omega)$ and the measure of  $\Omega_A$ goes to zero as $A\to\infty$. 

The term $g_1\equiv T(a_B u)$ can be estimated similarly to what is done in Ref.~\cite{C} ($g_1$ corresponds to the term $F^2_A(x)$ of that reference), that is,
\be
\label{g1p}
\Vert  T(a_B u)\Vert_p \leq C_1\Vert T(a_Bu)\Vert_{\frac{pd}{2p+d};2} \leq C_2 \Vert a_Bu\Vert_{\frac{pd}{2p+d}}\leq C_3\Vert u\Vert_{\frac{pd}{2p+d}}\leq C_4 \Vert u\Vert_{2;1}<\infty\,,\ee
$\forall p\in (\frac{d}{d-2},\infty)$ 
 if $3\leq d\leq 6$. In the string of estimates above, besides the arguments used to obtain \eqref{contractp},  we have used the fact that $a_B$ is bounded  and that the condition $\frac{1}{d} + \frac{2p +d}{pd} -\frac{1}{2}\geq  0$, yielding the Sobolev embedding $H_1(\Omega)\subset L^{\frac{pd}{2p+d}}(\Omega)$, is satisfied $\forall p\in (1,\infty)$ in $3\leq d\leq 6$ dimensions. The analysis in Ref.~\cite{C} regarding the term $T(a_B u)$ continues to hold  in $d>6$ dimensions, which are of no interest for the present work. 
 
In $d>3$ dimensions, having assumed the weaker hypothesis $b(x)\in L^{\frac{2d}{d+1}} (\Omega)$,  the estimate of the term $g_2$ (corresponding to $F^1_A(x)$ of Ref.~\cite{C})  introduces a limitation on $p$ which is not present in Ref.~\cite{C}. Nevertheless, our assumption is still sufficient to yield an improvement at the first step of the bootstrapping. More precisely, 
\be\label{g2p}
\Vert  T(b)\Vert_p \leq C_1\Vert T(b)\Vert_{{\frac{pd}{2p+d}};2} \leq C_2 \Vert b\Vert_{\frac{pd}{2p+d}}\leq C_3\Vert b\Vert_{\frac{2d}{d+1}}\;\quad 
\forall p\in \left(\frac{d}{d-2}, \frac{2d}{d-3}\right]\,,\mbox{ if } d>3\,.
\ee
Note that the condition  
$p\in \left({d}/{(d-2)},{2d}/{(d-3)}\right]$ in $d>3$ dimensions is needed for the last inequality in \eqref{g2p}. 

An analogous estimate in $3$ dimensions, using $3p/(2p+3)< 3/2$ and $b(x)\in L^q(\Omega)$ $\forall q< 3/2$, yields
\be\label{Tb-3}
\Vert T(b)\Vert_p < \infty\,\quad  \forall p\in \left(1, \infty\right)\,,\mbox{ if } d=3\,.
\ee  
In conclusion, under milder assumptions (than the ones stated in Theorem 3.2.1 in Ref.~\cite{C}) we prove $u\in L^p(\Omega)$ $\forall p\in (1,\infty)$ in $3$ dimensions, and $u\in L^{{2d}/{d-3}}(\Omega)$ in $d>3$ dimensions, over  a bounded domain $\Omega$. In particular,   $u\in L^8(\Omega)$ in $4$ dimensions.  
\qed

Note that, although no further improvement can be obtained in the general case (that is, when the term $b$ is prescribed with assigned fixed regularity), one may be able to take subsequent steps and obtain more regularity in all those cases in which there is feed-back into the term $b(x)$ coming from the solution. Fortunately, this is the case for $b\equiv h_1$, as in eq. \eqref{Dirichlet} obtained from doubling across the boundary of $\nst$, and  analogous contexts often present themselves in mathematical applications.   

Note also that a yet milder requirement  on the term $b$ of Theorem 3.3.2 in Ref.~\cite{C},  still sufficient to guarantee a first improvement in the regularity of $u$, is the condition $b(x)\in L^{2d/(d-1)}_{ -1}(\Omega)$, for $\Omega$ a bounded domain. For further details, as well as variants of Lemma \ref{L-RL}, see Ref.~\cite{M-reg}. 
 
\bn Returning to our problem, the previous analysis yields the following results. 

In $3$ dimensions,  the solution 
$\alpha_1\check\Phi$ to system \eqref{lin3} is in  $L^p(B_{{\bf a}; R})$ $\forall p$; thus $\tilde\Phi\in L^p(\mathcal U_{{\bf a}; R_1})$ $\forall p$. Therefore, $\Phi\equiv \tilde \Phi +\hat\varphi\in L^p({\mathcal U_{{\bf a};R_1}})$   $\forall p$, in which we have used that $\hat\varphi$, solution to system \eqref{Dirichlet}, satisfies $\hat\varphi\in H_{3/2}(\R^-\times \R^2)\subset   L^p(\threest)$ $\forall p$.

In $4$ dimensions, $\alpha\check\Phi$ is in $L^8(B_{{\bf a}; R})$;  thus $\tilde\Phi\in L^8(\mathcal U_{{\bf a}; R_1})$. Therefore $\Phi\equiv \tilde \Phi +\hat\varphi$ also belongs to $L^8(\mathcal U_{{\bf a}; R_1})$, in which we have used that $\hat\varphi\in H_{{3}/{2}}(\st)\subset L^8(\st)$. 

\sn \emph {Higher order regularity up to and including the boundary in $3$ dimensions.}  
After the first improvement, one takes the second summand of the first equation in \eqref  {wnlP}  to the right hand side of that equation, thus obtaining 
\be\label{step2-0} 
-\Delta \Phi = -{\mathcal P}^\prime(\Phi) \equiv f_2\mbox{ on } \threest\,.
\ee
 By replacing $\Phi$ with  $\tilde\Phi \equiv \Phi-\hat\varphi$ on the left hand side of eq. \eqref{step2-0}, one obtains 
 \be\label{higher3}
 -\Delta\tilde \Phi = -{\mathcal P}^\prime(\Phi) \equiv f_2\mbox{ on } \threest\,,
\ee
in which we have used the fact that $\hat \varphi$ satisfies the Laplace equation. Reflecting  the domain across the boundary, taking $\check \Phi$, $\check{f}_2$ to be the odd extensions of the functions $\tilde\Phi$ and   
$f_2$, respectively, and introducing a new smooth cut-off function $\alpha_2$, supported on the disk $B_{{\bf a}; R_1}$, such that $\alpha_2\equiv 1$ on $B_{{\bf a}; R_2}$, with $0<R_2<R_1$, one obtains 
\be\label{step2}
-\Delta (\alpha_2\check\Phi) = \alpha_2\check f_2 + \mathcal L_{\alpha_2}(\check\Phi)\quad\mbox{ on } \R^{3}\,,
\ee
in which $\check\Phi\in H_1(\R^3)$, and $\alpha_2\check\Phi\in H_{1; 0}(\Omega)$. 
The right hand side of eq. \eqref{step2}  is then in $L^2(\R^{3})$ (because $\Phi$, thus also $\mathcal P^\prime(\Phi)$ have been proved to belong to  $L^p(B_{{\bf a}; R_1})$ for all $p$, ${\mathcal L}_{\alpha_2}$ is a lower order differential operator, and $\alpha_2$ is smooth and compactly supported in $B_{{\bf a}; R_1}$). Thus $\check\Phi\in H_2(B_{{\bf a}; R_2})$ and 
$\tilde\Phi\in H_2(\mathcal U_{{\bf a}; R_2})$. By the Sobolev Lemma, $\tilde\Phi$ can be modified on a set of measure zero to yield 
$\tilde\Phi\in \mathcal C^0(\mathcal U_{{\bf a}; R_2})$ and, by the arbitrariness of the point $\bf a$, $\tilde\Phi\in \mathcal C^0((-\infty, 0]\times \R^2)$.

The term $\mathcal L_{\alpha_2}(\check\Phi)$ in eq. \eqref{step2} is now in $H_1(\R^3)\subset L^6(\R^3)$, thus the right hand side of that equation is now in $L^6(B_{{\bf a}; R_2})\;\forall p$,  yielding $\tilde\Phi\in L^6_2(\mathcal U_{{\bf a}; R_2})$. Eventually, one would obtain  $\check\Phi\in L^p_2(B_{{\bf a}; \bar R})\;\forall p$, $\tilde\Phi\in L^p_2(\mathcal U_{{\bf a}; \bar R})\;\forall p$ (for $0<\bar R<R_2$). Thus, $\tilde\Phi\in \mathcal C^1((-\infty, 0]\times \R^2)$ (using again 
the arbitrariness of the point $\bf a$).

This does not improve the regularity of $\Phi$, as the latter depends on the regularity of  $\hat \varphi$. 
Nonetheless, 
if the boundary value satisfies $\varphi\in\mathcal C^\infty(\threetzero)$, by differentiating  eq. \eqref{higher3} first with respect to tangential derivatives (that is, derivatives \emph{not} containing $\partial/\partial x^0$), then relating those to normal derivatives via eq. \eqref{higher3}, the same technique used above would eventually yield $\tilde\Phi$ and  $\Phi$ in $\mathcal C^\infty((-\infty,0]\times \R^2)$.

\sn \emph {Higher order regularity up to and including the boundary in $4$ dimensions.} 
We recall that, by Lemma \ref{L-RL} in $4$ dimensions, the solution $\alpha\check\Phi$ to system \eqref{lin3} is in $L^8(B_{{\bf a}; R})$, and that $\hat\varphi\in H_{{3}/{2}}(\st)\subset L^8(\st)$; thus $\tilde\Phi\in L^8(\mathcal U_{{\bf a}; R_1}), \Phi\equiv \tilde \Phi +\hat\varphi\in L^8(\mathcal U_{{\bf a}; R_1})$. 
After the first improvement, one takes the second summand of the first equation in \eqref  {wnlP}  to the right hand side of that equation and follows the procedure described for the $3$ dimensional case leading up to  equations analogous to \eqref{step2-0}, \eqref{higher3}, \eqref{step2}. The term  
$\alpha_2\check f_2$ in eq. \eqref{step2} 
 is now in $L^{\frac{8}{3}}(\R^4)$; one then eventually obtains  
$\check\Phi, \tilde\Phi\in L^{\frac{8}{3}}_2(\mathcal U_{{\bf a}; \bar R})$, for $0<\bar R<R_1$. Thus, by  Sobolev Lemma and the arbitrariness of  the boundary point ${\bf a}$, $\tilde\Phi\in \mathcal C^0 \left([0,\infty)\times\R^3\right)$. If the boundary value satisfies in addition $\varphi\in \mathcal C^\infty(\tzero)$, by following the same procedure outlined for the three-dimensional case,  one obtains $\tilde\Phi, \Phi\equiv \hat\varphi + \tilde\Phi\in\mathcal C^\infty((-\infty, 0]\times\R^3)$.

\sn\emph{Global control.}  Lemma \ref{L-harmonic} gives a global estimate $\hat\varphi\in H_{3/2}(\nst)$ over the unbounded domain $\nst$ for $\hat\varphi$ a solution to the Laplace equation with boundary value $\varphi\in\bb$. We now prove that the conditions on the coefficients of the polynomial $\mathcal P(\Phi)$  are sufficient to yield an analogous result for the solution $\Phi$ to the Dirichlet problem \eqref{wnlP}  in dimension  $2\leq d\equiv n+1\leq 4$.
To this purpose, we first extend Lemma \ref{L-harmonic} as follows:
\begin{lemma} \label{L-harmonic1}
Let $\Phi_L$ satisfy the linearized problem
\be \label{l}\left( L\right) \quad\qquad\left\{\begin{array}{ll}
\Lambda_L u\equiv -\Delta u + 2a_2 u =0\quad&\mbox{on } \nst\\
\Phi\in  \ma ({\varphi})\,,
\end{array}\right.
\ee
with prescribed $\varphi\in\bb$. Then $\Phi_L\in H_{{3}/{2}}(\nst)\cap \mathcal C^\infty(\nst)$.
If, in addition, the boundary value $\varphi\in \mathcal C^\infty(\ntzero)$, then $\Phi_L\in H_{{3}/{2}}(\nst)\cap\mathcal C^\infty\left((-\infty, 0]\times \R^n\right)$, that is, $\Phi_L$ is smooth all the way up to and including the boundary. 
\end{lemma}
\emph{Proof of Lemma \ref{L-harmonic1}:}  The result follows immediately from Lemma  \ref{L-harmonic} after noting that $\Phi_L-\hat\varphi$ satisfies 
\be
\quad\qquad\left\{\begin{array}{ll}
\Lambda_L u\equiv -\Delta u  +2a_2 u=  -2a_2 \hat\varphi\quad&\mbox{ on } \;\nst\\
u\in\ma(0)\,;
\end{array}\right.
\ee
thus $\Phi_L$ is as regular as $\hat\varphi$. 
\qed 

\begin{lemma} \label{L-control}
 Let $\Phi$ be the unique solution  to the  nonlinear Dirichlet problem \eqref{wnlP}. If $2\leq d\equiv n+1\leq 4$,   $\Phi\in H_{3/2}(\nst)$.  
\end{lemma}
\emph{Proof of Lemma \ref{L-control}:} A complication that arises on an unbounded domain is due to the presence of lower bounds for $p$ for the embeddings \eqref{SET}. An equation such as eq. \eqref{lb} is not applicable to derive an estimate on an unbounded domain because its coefficients $g(\Phi)$ and $r(\Phi; \hat\varphi)$ are only \emph{locally} in some $L^p$ space; more precisely, the various terms of $g$ and $r$ fail to lie in $L^p(\nst)$ for any shared value $p$. For this reason,  we use $\Lambda_L $ in place of $-\Delta$, $\hat\varphi$, solution to Laplace's equation,  in place of $\Phi_L$,  and derive from eq. \eqref{wnlP} an equation for $\tilde\Phi\equiv \Phi -\Phi_L\in\ma(0)$, such as 
$$\Lambda_L u + g_1(\Phi;\Phi_L) u = r_1(\Phi;\Phi_L)\,,$$
in which  $g_1$, $r_1$ belong to suitable Sobolev spaces over the unbounded domain $\nst$.

We treat separately the cases $d\equiv n+1=4$, $d=3$, and $d=2$. 

\sn \emph{Case $d=4$.}
We rewrite eq. \eqref{wnlP} as 
\begin{equation}\label{d4}
\Lambda_L u + g_1(\Phi)u =0\,.
\ee  
with $g_1(\Phi)\in L^{d/2}(\nst)$; explicitly, $g_1(\Phi) = 3a_3\Phi + 4a_4\Phi^2$. 
The function $\tilde\Phi_1\equiv \Phi-\Phi_L$ then satisfies
\be\label{Lgr}
\quad\qquad\left\{\begin{array}{ll}
\Lambda_L u +g_1(\Phi)  u=  -g_1(\Phi) \Phi_L\equiv r_1(\Phi;\Phi_L) \quad&\mbox{ on } \;\nst\\
u\in\ma(0)\,,
\end{array}\right.
\ee
in which $r_1(\Phi;\Phi_L)\equiv -g_1(\Phi) \Phi_L\in L^{{2d}/{d+1}}$ (having applied H\"older's inequality to $g_1(\Phi)\in L^{d/2}(\nst)$ and $\Phi_L\in H_{{3}/{2}}(\nst)\subset 
 L^{2d/(d-3)}(\nst)$). 
Let $\overline {g_1(\Phi)}$ the even extension of $g_1(\Phi)$  across the boundary $\ntzero$, and $\check\Phi_1$, $\check r_1(\Phi;\Phi_L)$ be the odd extensions across the boundary of $\Phi -\Phi_L$,   $r_1(\Phi;\Phi_L)$, respectively.
Then, $\overline {g_1(\Phi)}\in L^{d/2}(\R^d)$, $\check r_1\in L^{2d/d+1}(\R^d)$, and $\check\Phi_1$ is a solution to the linear system
\be\label{control}
\quad\qquad\left\{\begin{array}{ll}
\Lambda_L u +\overline {g_1(\Phi)}  u=  \check r_1 \quad&\mbox{ on } \;\R^d\\
u\in H_1(\R^d)\,.
\end{array}\right.
\ee
In the spirit of the theorem on contracting operators, extended to unbounded domains, we proceed as follows. 
Let $T_1$ be the solution operator for 
\be\label{controlG}
\quad\qquad\left\{\begin{array}{ll}
\Lambda_L u \equiv -\Delta u + 2a_2  u=  f \quad&\mbox{ on } \;\R^d\\
u\in H_1(\R^d)\,,
\end{array}\right.
\ee
with $f$ prescribed in a suitable Sobolev space.  Such solution operator exists because system \eqref{controlG} admits a unique solution. Furthermore, $T_1\,:\, H_k(\R^d)\to H_{k+2}(\R^d)$. 
Applying $T_1$ to the equation in  \eqref{control}, one obtains  
\be
\label{integral}
u = T_1 (- \overline {g_1}  u) + T_1(\check r_1) \equiv T_1(a_A u) + T_1(a_B u) + T_1 (\check r_1)\,,
\ee
in which the functions $a_A$, $a_B$ are defined as in \eqref{aAB}, with the understanding that $\Omega_A$ is now modified to be 
$$\Omega_A\equiv \{{\bf x}\in \Omega \;:\;\vert a(x)\vert\geq A\,,\;\vert x\vert>A\}\,,$$
for $A$ a positive fixed constant. 
One has  $T_1(a_B u)\in H_3(\R^d)$, $T_1 (\check r_1)\in L^{2d/d+1}_2(\R^d)\subset H_{3/2}(\R^d)$ for the right hand side of eq. \eqref{integral}; so, we are only left  
to show that 
\be 
T_1(a_A \cdot)\,:\, H_{{3}/{2}}(\R^d) \to H_{{3}/{2}}(\R^d)
\ee
is a contracting operator. If $d\geq 4$, one has indeed 
\be\label{string}
\Vert  T_1(a_A v)\Vert_{2; {\frac{3}{2}}}\leq  \Vert  T_1(a_A v)\Vert_{{\frac{2d}{d+1}};2} \leq  \Vert  a_A v\Vert_{\frac{2d}{d+1}}\leq \Vert a_A\Vert_{\frac{d}{2}} \,\Vert v\Vert_{\frac{2d}{d-3}}\leq \frac{1}{2} \,\Vert v\Vert_{2;{\frac {3}{2}}}\,,
\ee
in which  we have used the fact that $\Vert a_A\Vert_{d/2}\to 0$ as $A\to\infty$, Sobolev embeddings on unbounded domains and H\"older's inequality. Notice that the inclusion expressed by the first inequality above is a borderline case 
Because $\check \Phi_1$ (which, we recall,  is now defined as the odd extension  of $\Phi -\Phi_L$ across the boundary of $\nst$) is solution to the integral equation \eqref{integral}, one has 
\be\label{martin}
\frac{1}{2} \Vert  \check \Phi_1 \Vert_{2; {\frac{3}{2}}} \leq \Vert  \check \Phi_1 \Vert_{2; {\frac{3}{2}}} - \Vert  T_1(a_A \check \Phi_1)\Vert_{2; {\frac{3}{2}}}\leq \Vert T_1(a_B u)\Vert_{2; {\frac{3}{2}}} + \Vert T_1 (\check r_1)\Vert_{2; {\frac{3}{2}}}<\infty\,.
\ee
Restricting to the half space, $\Phi -\Phi_L\in H_{3/2}(\nst)$; thus  the solution $\Phi$ to \eqref  {wnlP} also satisfies 
$\Phi\in H_{{3}/{2}} (\nst)$,  in dimension  $d\equiv n+1=4$.

\sn \emph{Case $d=3$.} Eq. \eqref{wnlP} can be rewritten as   
$$\Lambda_L \Phi +  6a_6 (\Phi^5-\Phi_L^5)  = -( 3a_3 \Phi^2 +4a_4\Phi^3+ 5a_5\Phi^4 + 6a_6  \Phi_L^5)\;$$
thus the function $\tilde\Phi_1\equiv \Phi-\Phi_L $ satisfies 
\be
\label{c1}
\Lambda_L  \tilde\Phi_1+  g_1 (\Phi; \Phi_L)\,\tilde\Phi_1  =  r_1(\Phi;\Phi_L)\,,
\ee
in which $g_1 (\Phi; \Phi_L)\in L^p(\st)$ $\forall p\in [1/2, 3/2]$,  being a sum of monomials of degree $4$ in the variables $\Phi$, $\Phi_L$, and $r_1(\Phi;\Phi_L)= -( 3a_3 \Phi^2 +4a_4\Phi^3+ 5a_5\Phi^4 + 6a_6  \Phi_L^5)\in\L^{3/2}(\st)$; for these estimates we have also used that $\Phi_L\in L^p(\st)$ $\forall p\geq 2$. 

 Thus $\check\Phi_1$, odd extension of $\Phi -\Phi_L$, satisfies 
\be\label{control3}
\quad\qquad\left\{\begin{array}{ll}
\Lambda_L u +\bar g_1  u=  \check r_1 \quad&\mbox{ on } \;\R^d\\
u\in H_1(\R^d)\,,
\end{array}\right.
\ee
in which $\bar g_1\in L^p(\st)$ $\forall p\in [1/2, 3/2]$,  $\check r\in L^{3/2}(\st)$, yielding $T_1 (\check r_1)\in L^{3/2}_2(\R^3)\subset H_{3/2}(\R^3)$ (here, following the notation already established,  $\bar g_1$ denotes the even extension of $g_1$, $\check r_1$ denotes the odd extension of $r$). The operator  
\be 
T_1(a_A \cdot)\,:\, H_{{3}/{2}}(\R^3) \to H_{{3}/{2}}(\R^3)
\ee
is a contracting operator, as can  be seen applying the following string of inequalities 
\begin{align}\label{string3}
&\Vert  T_1(a_A v)\Vert_{2; \frac{3}{2}}\leq  \Vert  T_1(a_A v)\Vert_{\frac{3}{2};2} \leq \Vert  a_A v\Vert_{\frac{3}{2}}
\leq \liminf_{n\to\infty} \Vert a_Av\Vert _{\frac{3n}{2(n+1)}}^{\frac{n}{n+1} }\leq
\liminf_{n\to\infty} \Vert a_A\Vert _{\frac{3}{2}}^{\frac{n}{(n+1)} }\,\Vert v\Vert_{\frac{3n}{2}}^{\frac{n}{(n+1)} }\notag\\
&\leq \liminf_{n\to\infty} \Vert a_A\Vert _{\frac{3}{2}}^{\frac{n}{n+1} }\,\Vert v\Vert_{2; \frac{3}{2} -\frac{2}{n}}^{\frac{n}{n+1} } = \Vert a_A\Vert_{\frac{3}{2}} \,\Vert v\Vert_{2;\frac{3}{2}}\leq \frac{1}{2} \,\Vert v\Vert_{2;\frac{3}{2}}\,,\notag\\
\end{align}
in which we have applied Fatous's lemma to the sequence of non-negative functions $|a_A v|^{\frac{3n}{2(n+1)}}$.  
Thus, proceeding as for the case $d=4$,  ${\check \Phi}_1\in H_{3/2}(\R^3)$, yielding $\Phi_1 = \tilde\Phi +\Phi_L\in  H_{3/2}(\st)$. 

\sn \emph{Case $d=2$.} The procedure below for the case $d=2$ applies essentially unchanged to all dimensions, as long as the exponent of the polynomial $\mathcal P$ is strictly less than the critical exponent. In fact, in all such cases  there is no need to carry out an additional first step involving freezing part of the Euler Lagrange equations,  neither to the purpose of obtaining a first improvement in the bootstrapping, nor to the purpose of establishing global control over the unbounded domain. 

 Let $\Phi$ be the unique solution to \eqref{wnlP} for prescribed boundary value $\varphi$. In dimension $d=2$ 
the polynomial $\mathcal P^\prime(\Phi)$ satisfies 
$\mathcal P^\prime(\Phi)\in L^p(\onest)\;\forall p\geq 2$.
The function $\tilde\Phi\equiv\Phi-\hat\varphi$ satisfies 
\be
\label{lb2}\quad\qquad\left\{\begin{array}{ll}
-\Delta u=- \mathcal P^\prime(\Phi) &\mbox{ on } \;\onest\\
u \in\ma(0)\,,
\end{array}\right.
\ee 
and its odd extension $\check\Phi$ is in $H_1(\R^2)$ and satisfies 
\be
-\Delta u=  \check{r}\mbox{ on } \;\R^2\,,
\ee
in which $\check r\in L^p(\R^2)$ $(\forall p\geq 2)$ is the odd extension of $-\mathcal P^\prime(\Phi)$ across the boundary of $\onest$. 
This yields $\tilde\Phi\in L^p_2(\R^2)$ $\forall p\geq 2$, thus $\Phi\equiv \tilde\Phi+\Phi_L\in H_{3/2}(\R^2)$.

This concludes the proof of Theorem \ref{T-reg1}.\qed

\mn  {\bf   Remark.}  In all dimensions considered,  with some minor changes in the procedure, one would have been able to obtain the first improvement by using any extension of $\varphi$ in $H_{{3}/{2}} \left([0,\infty)\times\R^3\right)$.   However,  after the first step and to the purpose of establishing globel control, we have relied on the regularity properties of   $\hat\varphi$, harmonic extension of $\varphi$, or $\Phi_L$, solution to the linearized problem \eqref{l} with prescribed boundary value $\varphi$.  

\section{The energy functional} 
\label{S-conserv}
  For a distribution $\Phi\in \ma(\varphi)$, we define the Euclidean signature energy density
\be
\label{density}
\varepsilon(\Phi) \equiv \frac{1}{2}{ \dot\Phi}^2 - \frac{1}{2}\vert\nabla^\prime\Phi\vert^2 -\mathcal P (\Phi)\quad a.e. \mbox{ on } \nst\,,
\ee
in which $\nabla^\prime$ is the gradient with respect to the spacial variables ${\bf x^\prime}\equiv (x^1, \dots, x^n)$ only; thus $\vert\nabla^\prime\Phi\vert^2\equiv\sum_{j=1}^{n} (\p_j\Phi)^2$ and we are making the identifications $x^0\equiv t$, $\partial_0  \equiv \partial /\partial x^0= \p/\p t $. 

Associated with the energy density \eqref{density} we define the integral
\be
\label{mathcal F}
\mathcal F [\Phi]\equiv
\int_\nst \varepsilon(\Phi) \bigl(t, {\bf x^\prime}\bigr) \, dt \,d {\bf x^\prime}\,.
\ee
Finiteness of the Euclidean action, together with the condition
\be\label{P-}
\int_{\R^-\times \R^n} \mathcal P(\Phi) \,dt\,d {\bf x^\prime}> -\infty\,,
\ee
both facts guaranteed by the hypothesis $\Phi\in H_1(\nst)$, yield the boundedness of the integral \eqref{mathcal F}. (Notice that, actually, \eqref{coerciv.cond.} implies $\int_{\R^-\times \R^n} \mathcal P(\Phi) \,dt\,d {\bf x^\prime}\geq 0$). 
In more detail,
\ba\label{ineq}
&-\infty<  - \mathcal F [\Phi] =\int_{\R^-\times \R^n} \left(-\frac{1}{2}(\partial_0\Phi)^2 +\frac{1}{2}\sum_{j=1}^{n} (\partial_j\Phi)^2 +\mathcal P(\Phi)\right) \, dt \,d {\bf x^\prime}\notag\\
&\leq
\int_{\R^-\times \R^n} \left(\frac{1}{2}(\partial_0\Phi)^2 +\frac{1}{2}\sum_{j=1}^{n} (\partial_j\Phi)^2 +\mathcal P(\Phi)\right) \, dt \,d{\bf x^\prime}
\equiv \ie[\Phi]<\infty\,,\notag\\
\end{align}
in which the middle inequality is immediate, the last inequality is the finiteness of the Euclidean action, and the first inequality to the left follows straightforwardly from the second and third inequalities combined with \eqref{P-}. (Notice, that \eqref{P-} could be replaced by $\int (\partial_0 \Phi)^2 <\infty$, for the purpose of obtaining the first inequality above).

As a consequence, by Fubini's theorem, the \emph {energy}  of a distribution $\Phi\in\ma(\varphi)$,
\be\label{energy}
e [\Phi] (t) \equiv\int_{\R^n}\varepsilon(\Phi)(t,{\bf x^\prime}) \,d{\bf x^\prime}\,,
\ee
 is finite, thus can be defined, for almost all $t$'s in  $\R^{-}$.

Furthermore, formally, if $\Phi$ is a minimizer of $\ie$ over the affine space $\ma(\varphi)$, one has
\ba
\label{calc}  
&\p_0\varepsilon (\Phi) = \dot\Phi\ddot\Phi -\nabla^\prime\Phi\cdot \p_0(\nabla^\prime\Phi) - \mathcal P^\prime(\Phi)\dot\Phi \notag\\
&= \dot\Phi\ddot\Phi -\nabla^\prime\Phi\cdot \nabla^\prime(\p_0\Phi)  - \mathcal P^\prime(\Phi)\dot\Phi + \varsigma\notag\\
&=\dot\Phi(\ddot\Phi +\Delta^\prime\Phi-\mathcal P^\prime(\Phi))  - \nabla^\prime \cdot (\dot\Phi \nabla^\prime\Phi) + \varsigma \notag\\
&=  - \nabla^\prime \cdot (\dot\Phi \nabla^\prime\Phi) + \varsigma\,, \notag\\
\end{align}
in which we have denoted by $\Delta^\prime$ the Laplace operator with respect to the spacial coordinates only,  the last inequality holds because $\Phi$ satisfies the Euler-Lagrange equations, and one has 
\be
\label{varsigma}
\varsigma\equiv - \nabla^\prime\Phi\cdot \p_0(\nabla^\prime\Phi) +\nabla^\prime\Phi\cdot \nabla^\prime(\p_0\Phi)=0\,,
\ee
whenever the operators $\p_0$ and $\nabla^\prime$ commute, thus, in particular on smooth functions  $\Phi$. Thus, 
$\varsigma\equiv 0$ on $\nst$ by the interior regularity proved in \S  \ref{S-ellreg}.

\subsection{Conservation of the energy}
We prove conservation of energy for a minimizer $\Phi$ of the action, with prescribed  boundary value $\varphi\in \bb$. By the regularity theory developed in \S \ref{S-ellreg},  $\Phi$  is smooth in the interior; it is also smooth up to and including the boundary in the case the boundary value is assumed to be smooth (\cf Theorem \ref{T-reg1}). 

From equations \eqref{calc}, and \eqref{varsigma} applied to smooth functions $\Phi$, and the Fundamental Theorem of Calculus applied to the smooth functions of time $\partial_0\varepsilon (\Phi)(\,\cdot\,,  {\bf x^\prime})$ for $ {\bf x^\prime}$ fixed,
\begin{equation}
\label{smoothenergy}
\varepsilon (\Phi) (t_2,{\bf x^\prime}) - \varepsilon (\Phi) (t_1,{\bf x^\prime})= \int_{t_1}^{t_2} \p_0\varepsilon (\Phi) (t, {\bf x^\prime})\,dt =- \int_{t_1}^{t_2}\nabla^\prime \cdot (\dot\Phi \nabla^\prime\Phi)\,dt\,,\quad -\infty< t_1< t_2<0\,.
\end{equation}
Note that the above formula holds for $-\infty< t_1< t_2\leq 0$, if $\varphi\in\bb$ is assumed to be smooth. 
Let $E\subset\R^-$ be a set of measure zero  such that $e[\Phi] ({t})$ is finite for all $t\in \R^- \cap E^c$ (here $E^c$ denotes the complement of $E$).  Integrating \eqref{smoothenergy} over  $\R^n$, one obtains
\begin{equation}
\label{smooth1}
 e[\Phi] ({t_2}) - e[\Phi]({t_1})= - \int_{\R^n} d{\bf x^\prime}\int_{t_1}^{t_2}\nabla^\prime \cdot (\dot\Phi \nabla^\prime\Phi)\,dt\,,\quad \forall t_1< t_2\,, \mbox{ with } t_1, t_2\in \R^- \cap E^c\,.
\end{equation}
The quantities on the left hand side of this formula are finite and,  rewriting the integrand of the right hand side as the $(n+1)$-dimensional divergence of the vector field 
$$v({\bf x})\equiv v(t, {\bf x^\prime})\equiv\left(0, \dot\Phi\,\p_1\Phi, \dot\Phi\,\p_2\Phi, \dots, \dot\Phi\,\p_n\Phi\right)(t, {\bf x^\prime})\,,$$ 
yields
$$
e[\Phi] ({t_2}) - e[\Phi]({t_1}) = - \int_{\R^n\times [t_1,t_2] } \nabla \cdot v( {\bf x})\,d{\bf x}\,,\quad \forall t_1< t_2\,, \mbox{ with } t_1, t_2\in \R^- \cap E^c\,;
$$  
thus, applying Green's theorem and the definition of improper integral, 
\begin{align}
\label{smooth2}
e[\Phi] ({t_2}) - e[\Phi]({t_1}) &= \int_{\R^n} \left(v( t_1, {\bf x^\prime})- v(t_2, {\bf x^\prime})\right)\cdot (1,{\bf 0}) \,d{\bf x^\prime}\notag\\
&
-\lim_{R\to\infty}\int_{\{\vert {\bf x^\prime}\vert = R\}\times [t_1, t_2]}
\dot\Phi \nabla^\prime\Phi\cdot \frac{{\bf x^\prime}}{\vert{\bf x^\prime}\vert}\,
d\sigma_{cyl}\,,\quad \forall t_1< t_2\,, \mbox{ with } t_1, t_2\in \R^- \cap E^c
\,,
\end{align}
where $d\sigma_{cyl}$ is the surface element on the cylinder $\{\vert {\bf x^\prime}\vert = R\}\times [t_1, t_2]$.
Notice that Green's formula can be invoked for any fixed R, that is, 
\begin{align}
\label{smoothR}
e_R[\Phi] ({t_2}) - e_R[\Phi]({t_1}) &= \int_{\{\vert {\bf x^\prime}\vert \leq R\}} \left(v( t_1, {\bf x^\prime})- v(t_2, {\bf x^\prime})\right)\cdot (1, {\bf 0}) \,d{\bf x^\prime}\notag\\
&
-\int_{\{\vert {\bf x^\prime}\vert = R\}\times [t_1, t_2]}
\dot\Phi \nabla^\prime\Phi\cdot \frac{{\bf x^\prime}}{\vert{\bf x^\prime}\vert}\,
d\sigma_{cyl}\,,\quad -\infty< t_1< t_2<0
\,,
\end{align}
because all the integrands are smooth in the interior  of $\nst$;
 thus the  extension of  Green's formula \eqref {smoothR} to improper integrals, namely, in the limit as $R$ tends to infinity,  holds for those $t_1<t_2$ for which the left hand side remains bounded 
(that is,  for $t_1, t_2\in \R^-\cap E^c$). The fact that the expression to the left hand side of formula \eqref{smooth2} is finite for such $t$'s, and that the first summand to the right hand side vanishes (because the integrand vanishes),   
give a posteriori the existence (and finiteness) of the only term remaining in the right hand side of \eqref{smooth2}, namely, the term containing the limit.
Suppose now that   $\alpha\equiv\lim_{R\to\infty}|\int_{\{\vert {\bf x^\prime}\vert = R\}\times [t_1, t_2]}
\dot\Phi \nabla^\prime\Phi\cdot {\bf x^\prime}/ \vert {\bf x^\prime}\vert\,\,d\sigma_{cyl}|$ satisfies 
$0< \alpha$. Then, for $R$ sufficiently large, say $R\geq M$, one has that $|\int_{\{\vert {\bf x^\prime}\vert = R\}\times [t_1, t_2]}
\dot\Phi \nabla^\prime\Phi\cdot {\bf x^\prime}/ \vert {\bf x^\prime}\vert\,\,d\sigma_{cyl}|
>{\alpha}/{2}$
and, by H\"older's inequality,   
$$
\infty = \int_{M}^{\infty}  \frac{\alpha}{2} \,dR< \int_\nst \vert \dot\Phi \nabla^\prime\Phi\vert \,d{\bf x}\leq \Vert \nabla\Phi\Vert^2_{L^2(\nst)}<\infty\,.$$
This is a contradiction. Thus $\alpha=0$ and 
$e_{t_2}(\Phi) = e_{t_1}(\Phi)$, for all times $t_1< t_2\in \R^-\cap E^c$, yielding
\be
\label{C-0}
\int_{\R^n} \varepsilon(\Phi)(t, {\bf x^\prime})\, d {\bf x^\prime}= C\quad a.e. \mbox{ on } \R^-\,,
\ee 
in which $C$ is a constant, 
whenever $\Phi$ is a minimizer of the action functional $\ie$ over the space $\ma(\varphi)$, with  boundary value $\varphi\in \bb$. 

On the other hand, the elliptic regularity results in \S \ref{S-ellreg} show that the action minimizers $\Phi$ are smooth in the interior and belong to $H_{{3}/{2}}(\nst)$ (\cf Theorem \ref{T-reg1}) and, by Sobolev embeddings, noting also that restriction to submanifolds $\{t\}\times\R^n$ entails losing precisely half a derivative, 
\be
\label{C}
\int_{\R^n} \varepsilon(\Phi)(t, {\bf x^\prime})\, d {\bf x^\prime}= C\quad \forall t\leq 0\,.
\ee 

\sn This concludes the proof of conservation of energy along the flow for distributions $\Phi$ which minimize the action over $\ma(\varphi)$ for a generally prescribed boundary value $\varphi\in \bb$. 

\subsection{Vanishing of the energy at $t=0$ for the $\mathcal P(\Phi)$ theory}
\label{S-vanish}
In the preceding subsection we have proved that, if $\Phi$ is a minimizer of the Euclidean action $\ie$ over $\ma(\varphi)$, then 
\be\label{energyt}
 e[\Phi](t)\equiv
\int_{\R^n} \left(\frac{1}{2}(\p_0\Phi)^2 -\frac{1}{2}\sum_{j=1}^{n} (\p_j\Phi)^2 -\mathcal P(\Phi)\right)(t, {\bf x^\prime}) \,d{\bf x^\prime}=C\,,
\ee
in which $C$ is a constant depending only on $\Phi$ and, ultimately, by uniqueness of the minimizer, only on the prescribed boundary value $\varphi\in\bb$. 
The inequalities  \eqref{ineq} then yield the vanishing of $e[\Phi](t)$,  since 
$$
-\infty< -\int_{\R^-} C\,dt =  -\int_{\R^-} e[\Phi](t)\,dt  
<\infty\,
$$ 
is satisfied  if and only if $ C=  0$.
\subsection{The variational derivative of $S\equiv S_{(0)}$ with respect to boundary data $\varphi\in\bb$ and the Hamilton-Jacobi equation}
\label{S-HJ}
 In \S \ref{S-IFT} below we will show that the functional $S$ is $\mathcal C^\infty$ by first showing, by means of an implicit function theorem between Banach spaces, that the solution $\Phi_\varphi$ of \eqref{wnlP}, unique minimizer of $\ie$ with boundary value $\varphi$,  depends smoothly on the latter. 
As a consequence,  $v(\lambda)$ defined as 
\[v(\lambda)\equiv \Phi_{\varphi +\lambda\psi} - \Phi_\varphi\,,
\]
is differentiable with respect to $\lambda$; here $\varphi\in \bb$, $\psi\in \bb\cap\mathcal C^\infty(\ntzero)$, $\Phi_{\varphi +\lambda\psi}$ is  the (unique) minimizer of $\ie$ with boundary value $\varphi +\lambda\psi$.
In the present subsection, we compute directly the variational derivative of $S$ with respect to the boundary data 
$\varphi\in\bb$ utilizing such differentiability property, thus defining 
\[v_1\equiv v^\prime(\lambda)_{|_{\lambda=0}}\,.\]
We then derive the Hamilton-Jacobi equation obeyed by $S$.

To that purpose, 
we calculate  
\ba\label{var-der}
&\int_{\R^n}\frac{\delta S}{\delta \varphi}\,\psi\,d{\bf x^\prime}\equiv DS[\varphi](\psi) \equiv \lim_{\lambda\to 0} \frac{S[\varphi +\lambda\psi] - S[\varphi]}{\lambda}=
\lim_{\lambda\to 0} \frac{\ie(\Phi_{\varphi +\lambda\psi}) - \ie(\Phi_\varphi)}{\lambda}=\notag\\
&\notag\\
&\int_\nst\left( \nabla\Phi_\varphi\cdot\nabla v_1  + \mathcal P^\prime (\Phi_\varphi) \,
v_1 \right)  \,d{\bf x} =\int_\nst\left(-\Delta \Phi_\varphi +\mathcal P^\prime (\Phi_\varphi) \right)\,v_1 \,d{\bf x} \,+\notag\\
&\notag\\
&+\int_\ntzero
\psi\,\nabla\Phi_\varphi\cdot (1, {\bf 0}) \,d{\bf x^\prime} + \lim_{R\to\infty} \int_\rinfty v_1 \,\nabla\Phi_\varphi \cdot
\frac{{\bf x}}{\Vert  {\bf x}\Vert}\,d\sigma_{cyl} \notag\\
&\notag\\
& = \int_\ntzero
\psi\,\nabla\Phi_\varphi\cdot (1, {\bf 0}) \,d{\bf x^\prime} \equiv \int_\ntzero
\psi\,\frac{\partial \Phi_\varphi}{\partial x^0} \,d{\bf x^\prime} \,,\notag\\
\end{align} 
in which 
 $d\sigma_{cyl}$ is the surface element on the cylinder 
$\rinfty$. In the calculation above we have used integration by parts,  the fact that ${v_1}_{|_{\ntzero}} =\psi$,  the Euler-Lagrange equations satisfied by $\Phi_\varphi$ and the vanishing of the limit appearing in the third line. In fact, the equalities above, together with differentiability of $S$ proved in \S \ref{S-IFT}, as well as finiteness of $\int_\ntzero
\psi\,\nabla\Phi_\varphi\cdot (1, {\bf 0}) \,d{\bf x^\prime}$ (note that $\psi\in H_{1/2}(\ntzero)$, and $\nabla\Phi_\varphi|_\ntzero\in L^2(\ntzero)$ from the results in \S \ref{S-ellreg}), yield existence and finiteness of such limit $\alpha$. Suppose now  $\alpha\equiv \lim_{R\to\infty} \vert\int_\rinfty v_1 \,\nabla\Phi_\varphi \cdot
\frac{{\bf x}}{\Vert  {\bf x}\Vert}\,d\sigma_{cyl}\vert >0$; then, for $M$ sufficiently large, say $R\geq M$, one has that $\vert\int_\rinfty v_1 \,\nabla\Phi_\varphi \cdot
\frac{{\bf x}}{\Vert  {\bf x}\Vert}\,d\sigma_{cyl}\vert
>{\alpha}/{2}$ and, by H\"older's inequality
\be
\infty=\int_M^\infty \frac{\alpha}{2} dR\leq\int_\nst \left| v_1 \,\nabla\Phi_\varphi\right|\,d{\bf x}
\leq\Vert v_1\Vert_{L^2(\nst)}\Vert \nabla\Phi_\varphi\Vert_{L^2(\nst)}<\infty\,,
\ee
because $v_1\in H_1(\nst)$ by the smooth dependence of $\Phi_\varphi$ from the boundary data proved in \S \ref{S-IFT}. 
The above is a contradiction, thus $\alpha=0$. 

Arbitrariness of $\psi\in \bb\cap\mathcal C^\infty(\ntzero)$  in \eqref{var-der} then yields  
\be\label{vardot}
\frac{\delta S}{\delta \varphi({\bf x^\prime})}= \frac{\partial \Phi_\varphi}{\partial x^0}(0, {\bf x^\prime})\equiv \dot \Phi_\varphi(0)\,.
\ee
The vanishing of the energy at $t=0$ combined with \eqref{vardot}  finally yield the Hamilton-Jacobi equation
\be\label{HJeq}
\int_\ntzero \left(\frac{1}{2}\frac{\delta S}{\delta\varphi( {\bf x^\prime}) }\,\frac{\delta S}{\delta\varphi( {\bf x^\prime}) } -\frac{1}{2}\nabla^\prime\varphi({\bf x^\prime})\cdot\nabla^\prime\varphi({\bf x^\prime}) - \mathcal P(\varphi({\bf x^\prime}) ) \right) \,d{\bf x^\prime}=0
\ee
for the functional $S$ defined in \eqref{S}. 

\section{$\mathcal C^\infty$ regularity of ${S}$}
\label{S-IFT}
  As already mentioned in the final paragraph of Section \ref{S-setting}, uniqueness of the absolute minimizer is not necessary to define the
functional ${S}$ via  \eqref{S}. Nonetheless, non uniqueness is an obstruction to its differentiability.   This is because the gradient of a solution to a Hamilton-Jacobi equation should produce the complementary momentum for the trajectory through a given point in configuration space. Thus, if the trajectory is not unique, the existence of the gradient at the chosen point is compromised.
A second obstruction would be the existence of non vanishing \emph {Jacobi fields}. That situation presents itself when one analyzes the smoothness of the length-squared functional of a Riemannian manifold. In that case the failure of minimizing geodesics between fixed endpoints to be unique or the existence of nontrivial Jacobi fields along such geodesics are precisely the obstructions to establishing global smoothness for the length-squared functional. In contexts such as the Yang-Mills theory, the lack of uniqueness for the absolute minimizer for the Euclidean action corresponds to a lack of `everywhere' differentiability of the analogue to the functional ${S}$. (For a more in depth discussion  on this topic \cf \S  VI of Refs.~\cite{MMM-QM} and \cite{MMM-YM}). % ask them to review
For the polynomial theories studied in the present work, these obstructions can be ruled out, and we are able to achieve our goal of showing smoothness of the functional \eqref{S} as an application of the
implicit function theorem between Banach spaces.

We first prove the following theorem of smooth dependence on the initial data, for a solution to the nonlinear system associated with the given polynomial theory. We give the proof under the additional hypothesis 
\be\label{strictpoint-convex}
\mathcal P^{\prime\prime}(z)> 0
\ee (in \S \ref{S-setting} we have previously assumed the weaker condition $\mathcal P^{\prime\prime}(z)\geq 0$). In dimension $d=4$, for example, this amounts only to further assuming the condition $8a_4a_2\neq 3a_3^2$ on the coefficients of $\mathcal P$. 
\bt\label{sd}
The solution $\Phi_\varphi$ to the nonlinear Dirichlet problem \eqref  {wnlP} 
depends smoothly on the boundary data
$\varphi\in\bb$; that is, 
\begin{equation}\begin{array}{cc}
& H_1(\ntzero)\to H_1(\nst)\\
&\varphi\mapsto \Phi_\varphi\\
\end{array}
\end{equation}
is a smooth map.
\et
\emph{Proof of Theorem \ref{sd}:}\ Let ${\varphi}\in\bb$ be fixed arbitrarily and let $\fiL$ be the solution to the linearized boundary
value problem $\left( L\right) $ defined in \eqref{l}. 
Such a solution  $\fiL$ to $\left( L\right) $ is uniquely
determined by its boundary value ${\varphi}$ prescribed on
$\ntzero$. (Explicitly, the difference $\psi$ between any two solutions solves the
boundary value problem $-\Delta \psi +2a_2\psi=0$, $\psi \in
\ma(0)$; thus,  integration by parts against $\psi$ yields
$\int_\nst\left( \vert\nabla\psi\vert^2 +
2a_2\psi^2\right) \,d{\bf x} = 0$ -- the boundary terms vanish using the same arguments as in the calculation of the variation of $\ie$. Thus, $\psi \equiv  0$).

Following the notation already established, for given $\varphi\in   \mathcal B$, $\Phi_\varphi$ denotes the unique minimizer of the functional $\ie$  over the space $\ma({\varphi})$, and, as previously showed, is also the unique solution to the nonlinear Dirichlet problem given by eq. \eqref{EL} with boundary data $\varphi$ (\cf concluding remark of \S \ref{S-EL}). 

We now define the functional
\begin{align} \me\,:\, &\bb\times \ma(0)\to H_{-1}(\nst)\notag\\
& (\varphi, h)\to \Lambda \left( \fiL + h\right) \,,\notag\\
\end{align}
in which $\Lambda\equiv -\Delta +\pp^\prime (\cdot)$ is the nonlinear operator defined in \eqref{Lambda} (with $\mathcal Q\equiv\mathcal P^\prime$).
Since the map $\varphi\in\bb\to \fiL\in H_1 (\nst)$ is $\mathcal C^\infty$, so is $\me (\varphi, h)$ (with respect to both variables).

Let us now fix $\varphi_0\in\bb$. We denote by $h_0$ the difference between the solution $\fio\in\ma(\varphi_0)$ to eq. \eqref{EL} and the solution to the linear problem \eqref{l}, that is, 
\be h_0\equiv \Phi_{{\varphi}_0}
-{\Phi}_ {{\varphi}_0}^L\,,\ee
 One has $h_0\in\ma(0)$ and
\be \me ({\varphi}_0, h_0) = \Lambda (\fioL +
h_0) = \Lambda (\fio) =0\,,\ee
since $\fio$ satisfies the Euler-Lagrange equations \eqref{EL}.

Linearizing $\me$ at
$({\varphi}_0, h_0)$ with respect to the second variable $h$, one derives the linear operator
\ba
& D_2\me_0\equiv D_h \me{({\varphi}_0, h_0)} \,:\, T_{h_0}\ma(0)\simeq\ma(0) \to T_0\,H_{-1}(\nst)\simeq H_{-1}(\nst) \,,\notag\\
&D_2\me_0(\xi)\equiv \lim_{\lambda\to 0}\frac{\me({\varphi}_0, h_0 +
\lambda\xi) - \me({\varphi}_0, h_0)}{\lambda} =\lim_{\lambda\to
0}\frac{\Lambda(\fioL + h_0 + \lambda \xi)}{\lambda} \notag\\
&= \lim_{\lambda\to
0}\frac{\Lambda(\fio + \lambda \xi)}{\lambda}
= -\Delta \xi+ \mathcal P^{\prime\prime}(\fio)\,
\xi \,.
\end{align}
Here ${D_2\me_0}(\xi)$ is to be interpreted as an operator via the formula
\be
\label{def D2xi}
{D_2\me_0}(\xi)\,:\, f\in \ma(0)\mapsto \int_\nst f \,D_2\me_0 (\xi)\,d{\bf x} = \int_\nst \nabla f \cdot \nabla \xi \,d{\bf x}+ \int_\nst \mathcal P^{\prime\prime} (\fio)\,f\,\xi \, d{\bf x}\,,
\ee
in which the equality holds by integration by parts if $\xi$ has sufficient regularity and is otherwise to be understood as a definition of the left hand side for general $f \in \ma(0)$. 
Notice that the function $\mathcal P^{\prime\prime}(\fio)$ appearing in \eqref{def D2xi} depends on the chosen boundary data ${\varphi}_0\in \bb$ and that under the convexity and coerciveness 
conditions \eqref{strictpoint-convex} and \eqref{a2} imposed on the coefficients of $\mathcal P$,   
the linear operator
$D_2\me_0$ is a bicontinuous vector space isomorphism between
$\ma(0)$ and $H_{-1}(\nst)$.
To see this, begin by observing that the inner product on $H_1(\nst)$ given by 
\be
\label{ipa}
\langle f, g\rangle_{\mathcal P} \equiv \int_\nst \nabla f \cdot \nabla g \,d{\bf x}+ \int_\nst \mathcal P^{\prime\prime} (\fio)\,f\,g \, d{\bf x}
\ee and the usual inner product on $H_1(\nst)$ yield \emph {equivalent} norms, that is, induce the same topology. 
Directly, in the case of the standard $\Phi^4$ theory ($a_2\equiv1/2\,m^2$, $a_3=0$, $a_4\equiv\lambda>0$), using the notation $\Vert f \Vert^2_{\mathcal P} \equiv \langle f, f \rangle_{\mathcal P} $, one has $\pp^{\prime\prime} (\fio)= m^2 + 12\lambda\,\fio^2$ and 
\ba
\label{equiv1}
&\Vert f \Vert^2_{\mathcal P}\leq \int_\st \vert\nabla f \vert^2 \,d{\bf x}+ m^2\int_\st \vert f\vert^2\, d{\bf x}+12\lambda\left(\int_\st \fio^4\right)^{1/2}\left(\int_\st \vert f\vert^4\right)^{1/2}\notag\\
&\leq C_0\,\Vert f \Vert^2_{H_1(\st)} +12\lambda\, \Vert \fio\Vert^2_{L^4(\st)}\,\Vert f\Vert^2_{L^4(\st)} \leq C_1\,\Vert f\Vert^2_{H_1(\st)}\,,
\end{align}
in which,  having defined $\Vert f\Vert^2_{H_1(\st)}\equiv \int_\st \left(\vert\nabla f\vert^2+ (m^*)^2 f^2\right)\,d\bf x$ (notice that $m$ and $m^*$ have units 1/length), $C_0\equiv \max \,\{1, m^2/(m^*)^2\}$,   $C_1 \equiv C_0 + 12\lambda \,\Vert  \fio\Vert^2_{L^4(\st)}C^2$, with $C$ the Sobolev constant of the embedding $H_1(\st)\subset L^4(\st)$; the reverse inequality, namely, 
\be
\label{equiv2}
\Vert f\Vert^2_{H_1(\st)}\leq C_2 \Vert f \Vert^2_{\mathcal P}\,,
\ee
holds with $C_2 \equiv \max\,\{1, (m^*)^2/m^2\}$, for which we have used $m\neq 0$.
More in general, in dimension $d\equiv n+1 =2, 3$,  or $4$, for $\pp$ satisfying the hypotheses given in Secs. \ref{S-setting}, \ref{S-E!},  the estimate 
\be\label{equiv1g}
\Vert f \Vert^2_{\mathcal P}\leq C_1\,\Vert f\Vert^2_{H_1(\nst)}
\ee
is guaranteed by 
\ba\notag
& 0\leq \int_\nst \pp^{\prime\prime}(\fio) \,\vert f\vert^2\, d{\bf x} = \sum_{j=2}^k j(j-1)\int_\nst a_j\fio^{j -2}\,\vert f\vert^2 \, d{\bf x} \notag\\
&\leq \sum_{j=2}^k j(j-1)|a_j|\Vert \fio\Vert^2_{L^{2j-4}(\nst)}\,\Vert f\Vert^2_{L^4(\nst)} = 
C_1\,\Vert f\Vert^2_{H_1(\nst)}\,,\notag\\
\end{align}
while the reverse estimate 
\be
\label{equiv2g}
\Vert f\Vert^2_{H_1(\nst)}\leq C_2 \Vert f \Vert^2_{\mathcal P}
\ee
is guaranteed by the inequality
\be\label{suffices}
\int_\nst \mathcal P^{\prime\prime} (\fio)\,\vert f\vert^2 \, d{\bf x}\geq \alpha \int_\nst \vert f\vert^2 \, d{\bf x}
\ee
for some positive constant $\alpha$.  The latter is satisfied because the condition $\pp^{\prime\prime}(z)>0$ implies that the polynomial $\pp^{\prime\prime}(z)$ be bounded away from $0$. In dimension $d=4$, for example, $\alpha$ can be taken equal to $2a_2 -3a_3^2/4a_4$,  the $y$-coordinate of the vertex of the parabola $y=\pp^{\prime\prime}(z)$.

\sn 
The equivalence between the usual $H_1$-norm and $\Vert \cdot\Vert_{\mathcal P}$ proved above gives in particular that  $D_2\me_0 (\xi)$ is a bounded linear operator on $\ma(0)$.
(By explicit calculation,
\ba
&\Vert D_2\me_0 (\xi)\Vert_{H_1\to\R}\equiv \sup_{f\in \ma(0)} \frac{\vert \langle f \,,D_2\me_0 (\xi) \rangle_{L^2}\vert}{\Vert f\Vert_{H_1}}=  \sup_{f\in \ma(0)} \langle f, \xi\rangle_\pp\, \Vert f\Vert^{-1}_{H_1}\leq\notag\\
& \sup_{f\in \ma(0)} \left( \int_\nst \vert\nabla \xi \vert^2 +\mathcal P^{\prime\prime} (\fio)\,\vert \xi\vert^2\, d{\bf x}\right) ^{\frac{1}{2}}
\left( \int_\nst \vert\nabla f \vert^2 + \mathcal P^{\prime\prime} (\fio)\,\vert f \vert^2\, d{\bf x}\right) ^{\frac{1}{2}}\, \Vert f\Vert^{-1}_{H_1}\notag\\
&\leq C_1 \Vert \xi\Vert_{H_1}\,).
\end{align}
The linear operator $D_2\me_0$ is clearly one-to-one ($D_2\me_0 (\xi) =0 \implies \xi=0$ comes from \eqref{def D2xi} with $f=\xi$).
Moreover, because of the symmetry in the definition of $\langle f, g\rangle_{\mathcal P}$, one has $D_2\me_0(f)(g) = D_2\me_0(g)(f)$, which implies that $D_2\me_0\,:\, \ma(0)\to \ma(0)$ is a self-adjoint operator. 
Thus, $\left( Im \,D_2\me_0\right) ^\perp = Ker \, D_2\me_0^*= Ker \, D_2\me_0=0$.   Since $D_2\me_0^\ast$ is continuous, $Ker \,D_2\me_0^\ast$ is closed and $\ma(0)$ can be decomposed as  $\ma(0)= Ker\, D_2\me_0^\ast \oplus
\left( Ker\,D_2\me_0^\ast\right) ^\perp= Im \,D_2\me_0$, yielding surjectivity of $D_2\me_0$.

 This concludes the proof that  $D_2\me_0$,  obtained by differentiating $\me (\varphi, h)$ with respect to the second variable at the point $(\varphi_0, h_0)$, 
 is a bicontinuous isomorphism. The implicit function theorem between Banach spaces then states that  
there exist
neighborhoods $I\subset \bb$, $J\subset\ma(0)$ of ${\varphi}_0$, $h_0$ respectively, such that $\forall
\varphi\in I$ there exists a unique $h (\varphi)\in J$ for which $\me (\varphi,
h(\varphi))=0$ and that such map, $\varphi\in\bb\to h(\varphi)\in\ma(0)$, is $\mathcal C^\infty$. Because $\Phi_\varphi = \fiL+ h(\varphi)$, this concludes the proof of Theorem
\ref{sd}. \hfill$\Box$

At this point, using the fact that the functional $\ie\,:\, H_1(\nst)\to\R$ is smooth, the functional $S\,:\, \bb\to\R$ can be viewed as the composition of smooth maps as follows:
\be
S \,:\, \varphi\in\bb\to\Phi_\varphi\in \ma(\varphi)\to \ie[\Phi_\varphi]\in\R\,.
\ee
This yields finally the following theorem.
\bt
The functional \eqref{S} is smooth.
\et

\section{Decay of the approximate ground state wave functional for the polynomial theory $\mathcal P(\Phi)$}
\label{S-decay}
  We first prove a straightforward estimate of the type
\be\label{aim}
\vert\Omega_0(\varphi)\vert \leq \mathcal N \exp \left\{
\frac{    -\Vert \varphi\Vert^2_  {H_{\frac{1}{2}}(\R^n)}} {C}
\right\}\,,
\ee
in which $C$ is some constant (independent of $\varphi$) and $\Omega_0(\varphi)$ is defined in \eqref{Omega}.
A heuristic argument for a better estimate, based on a  so-called `virial argument',  which takes into account the presence of the higher order polynomial  term in $\ie$, is given in \S \ref{SS-virial} below.

\sn By the Trace Theorem,
\be\label{b-est}
\Vert \varphi\Vert_{H_{\frac{1}{2}}(\R^n)}\leq C_1 \Vert \Phi_\varphi\Vert_{H_1(\nst)}\,.
\ee
In order to obtain \eqref{aim}, it suffices to apply coerciveness of $\ie$, that is, an estimate of the type 
\be\label{aim2}
\Vert \Phi_\varphi\Vert^2_{H_1(\nst)}\leq C_2\, \ie[\Phi_\varphi]\,,
\ee
which is guaranteed by \eqref{coerciv.cond.} (which, we recall, follows from our assumptions \eqref{point-convex} and \eqref{a2} on the coefficients of $\mathcal P$; \cf \S \ref{SS-properties}). 
In fact, \eqref{b-est} combined with \eqref{aim2} yields
$$\Vert \varphi\Vert^2_{H_{\frac{1}{2}}(\R^{n})}\leq C\,\ie (\Phi_\varphi)\equiv C\, {S}[\varphi]\,,$$
thus
$$\exp \{- {S}[\varphi]\}\leq \exp \left\{\frac{-\Vert \varphi\Vert^2_{H_{\frac{1}{2}}(\R^n)}}{C}\right\}\,,$$
with $C= C_1^2 C_2$.
The massless case for the $\Phi^4$ theory on $\st$ and analogous cases in which one relaxes the hypothesis $a_2>0$ cannot be treated this way because, as observed in \S \ref{SS-properties}, $\ie$ is not coercive in those instances. 

\subsection{Virial estimates for $\mathcal P (\Phi)$ theories}
\label{SS-virial}
In the present section we give a heuristic argument for virial type estimates for $\mathcal P (\Phi)$ theories. For a rigorous argument one would also need to take into account the higher order corrections of the ground state functional, $S_{(1)}, S_{(2)} \dots$.

Here, we want to make a conjecture on the behavior of the ${S}_{(0)}$ functional under a (constant) rescaling of the form $\varphi\to \varphi_\lambda=  e^\lambda \varphi$. One cannot expect any simple behavior in general except in the limit of large $\varphi$. In order to establish such behavior in the limit, we consider the ratio

\be\label{ratio}
\mathcal R =\left.\frac{  d{S}_{(0)}[\varphi_\lambda]/d\lambda  }
                           {{S}_{(0)}[\varphi_\lambda]}\right| _{\lambda=0}
                        = \frac{\int_{\R^n}  \varphi(\mathbf x^\prime)\,\delta{S}_{(0)}[\varphi]/\delta  \varphi(\mathbf x^\prime)\,d\mathbf x^\prime}
                           {{S}_{(0)}[\varphi]} 
                           \ee 
and observe that in the case of \emph{free fields}, for which ${S}_{(0)}[\varphi ]$ is purely quadratic in  $\varphi$,  this ratio would simply be given by $\mathcal R = 2$ for all  $\varphi$.

	One expects both numerator and denominator to tend to infinity as $\Vert\varphi\Vert_{H_{1/2}(\R^n)}\to \infty$  and, indeed, this is true for the free field case and not hard to prove for the general case. As a side remark, an application of H\"older's inequality and the Hamilton-Jacobi equation obeyed by
	$\delta{S}_{(0)}[\varphi]/\delta  \varphi(\mathbf x^\prime)$ implies that the numerator of $\mathcal R(\varphi)$ cannot blow up until $\Vert\varphi\Vert_{H_1(\R^n)}$ does. We shall show in section \ref{SS-MFF} 
	below that, for arbitrary $\varphi$, ${S}_{(0)}[\varphi ]$  is bounded from below by a certain specific (massive) free field functional ${S}_{(0)}^{free} [\varphi ]$.  An explicit calculation of ${S}_{(0)}^{free} [\varphi ]$ can be found in Ref.~\cite{Maitra2007}. 
	
	  Assuming that  
	  $\lim_{\Vert\varphi\Vert_{H_{1/2}(\R^n)}\to \infty}\mathcal R(\varphi)$
	   exists, and using the fact that numerator and denominator of $\mathcal R$ blow up as $\Vert\varphi\Vert_{H_{1/2}(\R^n)}\to\infty$, we can appeal to L'Hospital's rule and evaluate such limit by differentiating numerator and denominator along  a curve $\varphi_t$ satisfying $\Vert\varphi_t\Vert_{H_{1/2}(\R^n)}\to\infty$ as $t$ approaches its limiting value $t^*$. It seems then to be convenient to take the limit along `solution curves' of the `gradient semi-flow' of the functional ${S}_{(0)}[\varphi ]$. Setting aside some subtleties, we calculate the formal time derivative of numerator and denominator along the flow by applying the functional differential operator	
	\be\label{ell}
\mathcal L =\int_{\R^n} \left(\frac{ \delta{S}_{(0)}[\varphi]}{\delta  \varphi(\mathbf y^\prime)}\right)
\frac{\delta}{\delta  \varphi(\mathbf y^\prime)}\,d\mathbf y^\prime 
                           \ee 
to each. One can simplify the result by appealing to the Hamilton-Jacobi equation satisfied by 
${S}_{(0)} [\varphi]$, namely, 
\be\label{hj}
\frac{1}{2}\int_{\R^n} \left(\frac{ \delta{S}_{(0)}[\varphi]}{\delta  \varphi(\mathbf z^\prime)}\right)
\left(\frac{\delta {S}_{(0)}[\varphi]}{\delta \varphi(\mathbf z^\prime)}\right)\,d \mathbf z^\prime 
  =      \int_{\R^n} \left(\frac{1}{2} \nabla^\prime   \varphi(\mathbf z^\prime)\cdot \nabla^\prime\varphi(\mathbf z^\prime)   +   \mathcal P( \varphi(\mathbf z^\prime))\right)\,d   \mathbf z^\prime  \,,         \ee      
                           in which the polynomial 
$$\mathcal P(\cdot)\equiv \sum_{j=2}^k a_j(\cdot)^j$$ of even degree $k$ satisfies the hypotheses illustrated in \S  \ref{S-setting}.   (We recall that  the coefficients of $\mathcal P$ are compatible with the assumed convexity and coerciveness of $\ie$, that $a_2$ and $a_k$  are strictly positive constants and that $k$ is an integer greater than $2$, but limited in terms of the dimension $n$ in order to not  exceed the critical exponent allowed by Sobolev embedding).

	The resulting formula for this ratio of `time' derivatives simplifies to
\be\label{timeder}
\mathcal T = 1 +
\frac 
{\int_{\R^n} \left(\nabla^\prime   \varphi(\mathbf y^\prime)\cdot \nabla^\prime\varphi(\mathbf y^\prime) + 2 a_2 \varphi^2(\mathbf y^\prime)+\dots + ka_k\varphi^k(\mathbf y^\prime)\right)\, d\mathbf y^\prime} 
{\int_{\R^n} \left( \nabla^\prime   \varphi(\mathbf z^\prime)\cdot \nabla^\prime\varphi(\mathbf z^\prime) + 2 a_2 \varphi^2(\mathbf z^\prime)+\dots + 2a_k\varphi^k(\mathbf z^\prime)\right)\, d\mathbf z^\prime}
\ee
Luckily the formula above is independent of ${S}_{(0)}(\varphi )$ and only depends upon the given `potential energy' from the Hamilton-Jacobi equation. This simplification is the main reason for proposing to compute the flow along the (Hamilton-Jacobi) solution curves instead of along the `rescaling curves'  $\varphi_\lambda=  e^\lambda \varphi$. Note also that in the absence of the higher order terms eq. \eqref{timeder} immediately reproduces the free field result  $\mathcal T\to 2$  without the need for taking $\varphi$  `large'.

	As already mentioned, in the spirit of the L'Hospital argument,  we replace $\varphi$   by $\varphi_t$, an arbitrary integral curve of the `gradient semi-flow' of the functional ${S}_{(0)}(\varphi )$, and attempt to evaluate the limit as $t$ increases to its limiting value $t^*$. 
	However,  if we flow along an  `integral curve' with norm 
	$\Vert\varphi_t (\cdot)\Vert_{H_{1/2}(\R^n)}\to\infty$   (as $t\to t^*$), then also $\Vert\varphi_t (\cdot)\Vert_{H_{1}(\R^n)}\to\infty$,  in which for sake of simplicity we define 
	\[\Vert\varphi (\cdot)\Vert^2_{H_1(\R^n)}\equiv \int_{\R^n} \left( \nabla^\prime   \varphi(\mathbf y^\prime)\cdot \nabla^\prime\varphi(\mathbf y^\prime) + 2 a_2 \varphi^2(\mathbf y^\prime)\right)
	\,d \mathbf y^\prime\,;\]
    we can then calculate the limit `universally' and arrive at the intuitively expected (see below) result, namely  $\mathcal T\to 1+k/2$.  To see this divide numerator and denominator on the right hand side of eq. \eqref{timeder} by the quantity 
    $\Vert\varphi (\cdot)\Vert^{k}_{H_1(\R^n)}$   and recall that for $p$ not exceeding the critical exponent for the chosen dimension $n$ one has   
    \be\label{crit1}
    \Vert\varphi(\cdot)\Vert_{L^p(\R^n)}    <      C(n, p) \Vert\varphi(\cdot)\Vert_{H_1(\R^n)}
    \ee 
    for some constant $C(n, p)$.  Each term in the numerator and denominator of \eqref{timeder}   except the top order ones, will then consist of the square of an $H_1(\R^n)$-norm or the $p$-th power of an $L^p(\R^n)$-norm divided by $\Vert\varphi (\cdot)\Vert^{k}_{H_1(\R^n)}$ with 
     $2 < p < k$. Thus each term except the top order ones will tend asymptotically to $0$ as $\Vert\varphi (\cdot)\Vert_{H_1(\R^n)}\to\infty$,  leaving the desired result $\mathcal T\to 1+k/2$.
     We have referred to this as the `intuitively expected' result since, from the form of the Hamilton-Jacobi equation satisfied by ${S}_{(0)} [\varphi ]$, we seem to need ${S}_{(0)} [\varphi]$ scaling like $\varphi^{k + 1}$ for large  $\varphi$   in order to match the behavior of the given potential energy for large arguments.

\subsection{Comparison with a (massive) free field}
\label{SS-MFF}
Recall that  we defined ${S}_{(0)} [\varphi ]$, for arbitrary fixed boundary data  $\varphi$  (lying in the appropriate trace space), to be the value of the Euclidean signature action functional
\[\mathcal I_{es} [\Phi] \equiv \int_{\R^n} \int_{-\infty}^0\left( 1/2\, \dot\Phi^2  + 1/2\, \nabla^\prime\Phi\cdot\nabla^\prime\Phi + \mathcal P(\Phi)\right) \,dt\,d{\bf x^\prime}\,,\]
	           evaluated on the minimizer $\Phi_\varphi$ having the chosen boundary data, i.e.,                                              
\[{S}_{(0)}[\varphi] = \mathcal I_{es} [\Phi_\varphi] \,.\]
However, we also required that the polynomial $\mathcal P$  satisfy a coerciveness condition
$\mathcal P(\Phi) \geq C\Phi^2$, 
for some constant $C > 0$. Write for convenience $C =1/2\, m_0^2$,  for some positive `mass' $m_0$, and note that if $\mathcal P(\Phi)$ contains only even terms with positive coefficients one could simply take  $1/2\,m_0^2 = a_2$. In any case it follows from the assumed inequality,
\[\mathcal P(\Phi) \geq 1/2 \,m_0^2\Phi^2\,,\]
that 
\begin{align}\label{free}
{S}_{(0)}[\varphi] &= \mathcal I_{es} [\Phi_\varphi] \equiv
\int_{\R^n} \int_{-\infty}^0\left( 1/2 \,\dot\Phi_\varphi^2  + 1/2 \,\nabla^\prime\Phi_\varphi\cdot\nabla^\prime\Phi_\varphi + \mathcal P(\Phi_\varphi)\right) \,dt\,d{\bf x^\prime}\notag\\
&\geq \mathcal I_{es}^{free} [\Phi_\varphi] \equiv
\int_{\R^n} \int_{-\infty}^0\left( 1/2 \,\dot\Phi^2  + 1/2\, \nabla^\prime\Phi_\varphi\cdot\nabla^\prime\Phi_\varphi +1/2\, m_0^2\,\Phi^2_\varphi \right) \,dt\,d{\bf x^\prime}\notag\\
&\geq \mathcal I_{es}^{free} \,[\Phi^{free}_\varphi] \equiv {S}_{(0)}^{free}\,[\varphi] \,,\notag\\
\end{align} 
in which $\Phi^{free}_\varphi$ is the minimizer of 
\[\mathcal I_{es}^{free} \,[\Phi_\varphi] \equiv
\int_{\R^n} \int_{-\infty}^0\left( 1/2 \,\dot\Phi^2  + 1/2\, \nabla^\prime\Phi_\varphi\cdot\nabla^\prime\Phi_\varphi +1/2\, m_0^2\,\Phi^2_\varphi \right) \,dt\,d{\bf x^\prime}\] 
having the same fixed boundary data  $\varphi$  as that chosen for ${S}_{(0)}$.  Thus,
\[{S}_{(0)}[\varphi] \geq {S}_{(0)}^{free}\,[\varphi] \]
fore every $\varphi$   in the designated trace space. 
In particular this shows (see the beginning of \S  \ref{S-decay})  that  
$\exp \{- {S}_{(0)}[\varphi]/\hbar\}$ decays at least as rapidly as some specific Gaussian. Recall however that whereas in the true free field case the (ground state) quantum corrections to 
${S}_{(0)}^{free}\,[\varphi]$  all vanish, this is not generally true for ${S}_{(0)}[\varphi]$.

\mn
\emph{\bf Remark}. There are some subtleties in the argument presented in \S  \ref{SS-virial} which we have not yet discussed.  
In particular, a difficulty in making the above argument rigorous arises through the fact that extension of the formal `integral curves' of the `gradient semi-flow' of the functional ${S}_{(0)}[\varphi]$
to positive $t$'s (for data specified at $t = 0$), does not necessarily make sense. Using our regularity results (\cf \S  \ref{S-ellreg}) there is a clear mathematical sense to such `curves' for $t < 0$ but, to extend them in the opposite temporal direction seems problematic in general, especially when the boundary data chosen is as `rough' as possible. However, such rough data cannot arise at an interior point of such a hypothetically extendible curve. In fact, the global control and interior regularity established in  \S \ref{S-ellreg} for a solution $\Phi_\varphi$ to the nonlinear Dirichlet problem \eqref{wnlP}, ensure $\Phi_\varphi\in H_{3/2}(\nst)\cap \mathcal C^\infty(\nst)$. Thus, $\varphi_t\in H_{1}(\R^n)\cap \mathcal C^\infty(\R^n)\subset\bb$,  $\forall t<0$, and one could regard the smoothed interior data at some $t < 0$ as new `initial data' for a curve that is in fact extendible (at least back to the original $t = 0$ starting point) and 
make presumably precise sense of the argument for a dense subset of the full space of initial data. How best to accomplish this is under study.
\section{Conclusion}
\label{sec:conclusion}
A key feature of the current quantization program, when applied to finite-dimensional, \textit{harmonic} oscillators, is that it regenerates all the well-known, \textit{exact} results for both ground and excited states, correctly capturing not only the eigenvalues but the \textit{exact eigenfunctions} as well \cite{MMM-QM, DimassiSjostrand1999, Helffer1988}. One finds for example that the fundamental solution to the relevant (inverted-potential-vanishing-energy) Hamilton-Jacobi equation, for an \textit{n}-dimensional oscillator (with mass \textit{m} and (strictly positive) oscillation frequencies \(\lbrace\omega_i\rbrace\)) is given by
\begin{equation}\label{eq:901}
\mathcal{S}_{(0)}(\mathbf{x}) = \frac{1}{2} m \sum_{i=1}^n \omega_i(x^i)^2
\end{equation}
and that all higher order corrections to the ground state wave function vanish identically leaving the familiar gaussian
\begin{equation}\label{eq:902}
\zero{\Psi}(\mathbf{x}) = \zero{N}_\hbar\; e^{-\frac{m}{2\hbar}\sum_{i=1}^n\omega_i(x^i)^2}
\end{equation}
where \(\mathbf{x} = (x^1, \cdots , x^n)\) and \(\zero{N}_\hbar\) is a normalization constant.

The construction of excited states begins with the observation that the only globally regular solutions to the corresponding, leading order `transport equation' are comprised of the monomials
\begin{equation}\label{eq:903}
\emm{\varphi}_{(0)}(\mathbf{x}) = (x^1)^{m_1} (x^2)^{m_2} \cdots (x^n)^{m_n},
\end{equation}
where \(\mathbf{m} = (m_1, m_2, \cdots , m_n)\) is an \textit{n}-tuple of non-negative integers with \(|\mathbf{m}| := \sum_{i=1}^n m_i > 0\), and proceeds, after a finite number of unequivocal steps, to assemble the exact eigenstate prefactor
\begin{equation}\label{eq:904}
\emm{\varphi}_\hbar(\mathbf{x}) = \emm{N}_\hbar\; H_{m_1} \left(\sqrt{\frac{m\omega_1}{\hbar}}x^1\right) H_{m_2} \left(\sqrt{\frac{m\omega_2}{\hbar}}x^2\right) \cdots H_{m_n} \left(\sqrt{\frac{m\omega_n}{\hbar}}x^n\right)
\end{equation}
where \(H_k\) is the Hermite polynomial of order \textit{k} (and \(\emm{N}_\hbar\) is the corresponding normalization constant) \cite{MMM-QM, DimassiSjostrand1999, Helffer1988}.

Modulo some apparently quite modest technicalities, needed to handle a continuous range of frequencies, it seems clear that when these same  (Euclidean signature semi-classical) methods are applied to \textit{free}, bosonic field theories they will simply regenerate the well-known (Fock-space) exact solutions for these systems. Indeed, the fundamental solutions to the relevant (Euclidean signature) Hamilton-Jacobi equations are explicitly known for the most interesting cases (\cite{Maitra2007}, and from a different perspective \cite{WheelerGeometrodynamics}).

While there is nothing especially astonishing about being able to rederive such well-known, exact results in a different way, we invite the reader to compare them with those obtainable via the textbook WKB methods of the physics literature \cite{BrackBhaduri2008, Orozio1988}. Even for purely \textit{harmonic} oscillators conventional WKB methods yield only rather rough approximations to the wave functions and are, in any case, practically limited to one-dimensional problems and to those reducible to such through a separation of variables. The lesser known Einstein Brillouin Keller (or EBK) extension of traditional semi-classical methods does apply to higher (finite-) dimensional systems but only to those that are completely integrable at the classical level \cite{Stone, BrackBhaduri2008}. In sharp contrast to these well-established approximation methods the Euclidean signature semi-classical program that we are advocating here requires neither classical integratibility nor finite dimensionality for its implementation.

As was discussed in the concluding section of Ref.~\cite{MMM-QM} our fundamental solution $S_{(0)}(\mathbf{x})$, to the (inverted-potential-vanishing-energy) Hamilton-Jacobi equation for a coupled system of nonlinear oscillators has a natural geometric interpretation.  The graphs, in the associated phase space $T^*\R^n$, of its positive and negative gradients correspond precisely to the stable ($W^s(p) \subset T^*\R^n$) and unstable ($W^u(p) \subset T^*\R^n$) Lagrangian submanifolds of the assumed equilibrium point $p = ( \zero{\mathbf{x}}, \mathbf{0} ) \in T^*\R^n$:
\begin{align*}
W^u(p)  &= \left\{ \left( \mathbf{x}, \mathbf{p} \right) \;:\; \mathbf{x} \in \R^n \, , \ \mathbf{p} = \nabla S_{(0)} \left( \mathbf{x} \right) \right\} \notag \\
W^s(p)  &= \left\{ \left( \mathbf{x}, \mathbf{p} \right) \;:\; \mathbf{x} \in \R^n \, , \ \mathbf{p} = - \nabla S_{(0)} \left( \mathbf{x} \right) \right\}.
\end{align*}

Another interesting result established for the \textit{nonlinear} oscillators discussed in Ref.~\cite{MMM-QM} is that the first quantum, `loop correction', \(\mathcal{S}_{(1)}(x^1, \cdots , x^n)\), to the (`tree level') fundamental solution, \(\mathcal{S}_{(0)}(x^1, \cdots , x^n)\), also has a natural geometric interpretation in terms of `Sternberg coordinates' for the gradient (semi-) flow generated by this fundamental solution. Sternberg coordinates, by construction, linearize the Hamilton-Jacobi flow equations
\begin{equation}\label{eq:905}
m\frac{dx^i(t)}{dt} = \frac{\partial\mathcal{S}_{(0)}}{\partial x^i} \left( x^1(t), \cdots , x^n(t)\right)
\end{equation}
to the form
\begin{equation}\label{eq:906}
\frac{dy^i(t)}{dt} = \omega_iy^i(t)\quad \text{(no sum on \textit{i})}
\end{equation}
through, as was proven in Ref.~\cite{MMM-QM}, the application of a global diffeomorphism
\begin{align}
\mu : \mathbb{R}^n \longrightarrow \mu (\mathbb{R}^n) \subset \mathbb{R}^n = \left\lbrace (y^1, \cdots , y^n)\right\rbrace ,\label{eq:907}\\
\mathbf{x} \longmapsto \mu (\mathbf{x}) = \left\lbrace y^1(\mathbf{x}), \cdots , y^n(\mathbf{x})\right)\label{eq:908}
\end{align}
that maps \(\mathbb{R}^n\) to a star-shaped domain \(K = \mu (\mathbb{R}^n) \subset \mathbb{R}^n\) with \(\mu^{-1}(K) \approx \mathbb{R}^n = \left\lbrace (x^1, \cdots , x^n)\right\rbrace\).

Though not strictly needed for the constructions of Ref.~\cite{MMM-QM}, Sternberg coordinates have the natural feature of generating a Jacobian determinant for the Hilbert-space integration measure that \textit{exactly cancels} the contribution of the first quantum `loop correction', \(\mathcal{S}_{(1)}(\mathbf{x})\), to inner product calculations, taking, for example,
\begin{equation}\label{eq:909}
\begin{split}
\left\langle\emm{\Psi},\emm{\Psi}\right\rangle &:= \int_{\mathbb{R}^n} \left|\emm{\Psi}(\mathbf{x})\right|^2 d^nx\\
&= \int_{\mu (\mathbb{R}^n)} \left|\emm{\Psi} \circ \mu^{-1}(\mathbf{y})\right|^2 \sqrt{\det{g_{**}(\mathbf{y})}} d^ny
\end{split}
\end{equation}
to the form
\begin{equation}\label{eq:910}
\left\langle\emm{\Psi},\emm{\Psi}\right\rangle = \int_{\mu (\mathbb{R}^n)} \left|\left[\emm{\varphi}\; e^{\frac{-\mathcal{S}_{(0)}}{\hbar} - \frac{\hbar}{2!} \mathcal{S}_{(2)} \dots}\right] \circ \mu^{-1}(\mathbf{y})\right|^2 \sqrt{\det{g_{**}(\mathbf{0})}} d^ny
\end{equation}
where, in the last integral, the contribution of \(\mathcal{S}_{(1)} \circ \mu^{-1}(\mathbf{y})\) to the wave function
\begin{equation}\label{eq:911}
\emm{\Psi} \circ \mu^{-1}(\mathbf{y}) = \left[\emm{\varphi}\; e^{-\frac{\mathcal{S}_{(0)}}{\hbar} - \mathcal{S}_{(1)} - \frac{\hbar}{2!} \mathcal{S}_{(2)} \dots}\right] \circ \mu^{-1}(\mathbf{y})
\end{equation}
has precisely cancelled the non-Cartesian measure factor \(\sqrt{\det{g_{**}(\mathbf{y})}}\), leaving the constant (Euclidean) factor \(\sqrt{\det{g_{**}(\mathbf{0})}}\) in its place. Roughly speaking therefore, this role of \(\mathcal{S}_{(1)}\) is to `flatten out' the Sternberg coordinate volume element, reducing it to ordinary Lebesgue measure (albeit only over the star-shaped domain \(\mu (\mathbb{R}^n)\)), by exactly cancelling the Jacobian determinant that arises from the coordinate transformation.

For purely \textit{harmonic} oscillators the original (Cartesian) coordinates \(\lbrace x^1, \cdots , x^n\rbrace\) are already of Sternberg type, \(\mathcal{S}_1(\mathbf{x})\) accordingly vanishes and Hilbert space inner product integrals reduce to their familiar, Cartesian form. For \textit{free fields}, on the other hand, such formal, stand-alone `Lebesgue measures' are, of course, mathematically undefined but, when combined with the universally appearing convergence factors, \(N_\hbar\; e^{-\mathcal{S}_{(0)}[\varphi]/\hbar}\), arising in all of the associated wave functionals, can be interpreted as providing rigorously defined gaussian measures for Fock space \cite{GlimmJaffeBook}.

For the nonlinear oscillators discussed in Ref.~\cite{MMM-QM}, Sternberg coordinates also have the remarkable property of allowing the leading order transport equation for \textit{excited states} to be solved in closed form. Indeed, the regular solutions to this equation are comprised of the monomials
\begin{equation}
\emm{\varphi}_{(0)}(\mathbf{y}) = (y^1)^{m_1} (y^2)^{m_2} \cdots (y^n)^{m_n}
\end{equation}
wherein, precisely as for the harmonic case, the \(m_i\) are non-negative integers with \(|\mathbf{m}| := \sum_{i=1}^n m_i > 0\). On the other hand the higher order corrections, \(\left\lbrace\emm{\varphi}_{(k)}(\mathbf{y});\quad k = 1, 2, \cdots\right\rbrace\), to these excited state prefactors will not in general terminate at a finite order as they do for strictly \textit{harmonic} oscillators but they are nevertheless systematically computable through the sequential integration of a set of well-understood linear transport equations \cite{MMM-QM}. Formal expansions (in powers of \(\hbar\)) for the corresponding (ground and excited state) energy \textit{eigenvalues} are uniquely determined by the demand for global regularity of the associated eigenfunction expansions. More precisely one finds, upon integrating the relevant transport equation at a given order, that the only potential breakdown of smoothness for the solution, would necessarily occur at the `origin' \(\mathbf{x} = \mathbf{0}\) (chosen here to coincide with the minimum of the potential energy) but that this loss of regularity can always be uniquely avoided by an appropriate choice of eigenvalue coefficient at the corresponding order \cite{MMM-QM}.

To compute such higher order `loop' corrections for a field theoretic problem of the type discussed herein one will first need to regularize the formal functional Laplacian that arises in the Schr\"{o}dinger operator (II.1) and that will reoccur in each of the transport equations which result from substituting ans\"{a}tze such as (II.3), (II.5) and (II.6) into the time independent Schr\"{o}dinger equation (II.4) and expanding formally in powers of \(\hbar\). Solving these transport equations for the `loop corrections', \(\left\lbrace\mathcal{S}_{(1)} [\varphi(\cdot)], \mathcal{S}_{(2)} [\varphi (\cdot)], \cdots\right\rbrace\), to the ground state wave functional simply amounts to \textit{evaluating} sequentially computable, smooth functionals on the Euclidean action minimizers, \(\Phi_\varphi\), for arbitrary chosen boundary data, \(\varphi (\cdot)\).

Solving the transport equations for excited states is somewhat more involved since these equations entail a lower order term in the unknown but the technology for handling this, is well understood \cite{MMM-QM, DimassiSjostrand1999, Helffer1988}. If, for example, a Sternberg diffeomorphism could be shown to exist for field theoretic problems of the type discussed herein then the leading order, excited state transport equation could be solved in closed form. Otherwise though one could simply fall back on the machinery developed in Refs.~\cite{MMM-QM, DimassiSjostrand1999, Helffer1988}, which does not assume the existence of Sternberg coordinates, and solve this and the corresponding higher order excited state equations in a less direct fashion. In either case it is intriguing to note that the excited states for \textit{interacting} field theories would be naturally labeled by sequences of (integral) `particle excitation numbers' in much the same way that the Fock-space excited states of a free field are characterized.

One often hears that the fundamental particle interpretation of \textit{interacting} quantized fields hinges upon their approximation by corresponding \textit{free} fields. This is unsatisfactory at best since, of course, an elementary particle cannot `turn off' its self-interactions in order to conveniently behave, even asymptotically, like a Fock-space, free field quantum. As we have already emphasized one of the natural features of this (Euclidean signature-semi-classical) program is that it maintains the dynamical nonlinearities of an interacting quantum system intact at every level of the analysis rather than attempting to reinstate nonlinear effects gradually through a perturbative expansion.

The reader may well have noticed that our methods precisely apply, at least for bosonic field theories, to those cases which are perturbatively renormalizable and indeed we do not expect to be able to remove the needed regularization `cutoffs' without the corresponding necessity to absorb divergences into the associated `coupling' constants \(\lbrace a_j\rbrace\) of the system under study. The details of such a (Euclidean signature-semi-classical) renormalization program are currently the object of ongoing research.

\bibliographystyle{unsrt}   % this means that the order of reference is determined by the order in which the \cite and \nocite commands appear
\bibliography{MariniMaitraMoncrief}  % list here all the references that you need
			    
\ed